\documentclass[twocolumn]{aastex631}
\usepackage{CJK}
\usepackage{color, amssymb, amsmath, natbib, color}
\usepackage{multirow, placeins, subfigure}
\usepackage{graphicx}
\usepackage{epstopdf}
\usepackage{bm}
\usepackage{overpic}
\usepackage{hyperref}
\usepackage{natbib}

\newcommand{\msunh}{\>h^{-1}\rm M_\odot}
\newcommand{\msunhh}{\>h^{-2}\rm M_\odot}
\newcommand{\Lsunhh}{\,h^{-2}\rm L_\odot}

\newcommand{\mpch}{\>h^{-1}{\rm {Mpc}}}

\begin{document}
\begin{CJK*}{UTF8}{gbsn}

\title{Measuring the conditional luminosity and stellar mass functions of galaxies by combining the DESI LS DR9, SV3 and Y1 data}

\correspondingauthor{Xiaohu Yang, Yirong Wang}
\email{xyang@sjtu.edu.cn, wyirong@sjtu.edu.cn}

\author[0000-0003-3203-3299]{Yirong Wang (王艺蓉)}
\affiliation{Department of Astronomy, School of Physics and Astronomy, and Shanghai Key Laboratory for Particle Physics and Cosmology, Shanghai Jiao Tong University, Shanghai 200240, People's Republic of China}

\author[0000-0003-3997-4606]{Xiaohu Yang (杨小虎)}
\affiliation{Department of Astronomy, School of Physics and Astronomy, and Shanghai Key Laboratory for Particle Physics and Cosmology, Shanghai Jiao Tong University, Shanghai 200240, People's Republic of China}
\affiliation{Tsung-Dao Lee Institute, and Key Laboratory for Particle Physics, Astrophysics and Cosmology, Ministry of Education,
Shanghai Jiao Tong University, Shanghai 200240, People's Republic of China}

\author[0000-0002-5632-9345]{Yizhou Gu (顾一舟)}
\affiliation{Department of Astronomy, School of Physics and Astronomy, and Shanghai Key Laboratory for Particle Physics and Cosmology, Shanghai Jiao Tong University, Shanghai 200240, People's Republic of China}

\author{Xiaoju Xu (徐笑菊)}
\affiliation{Department of Astronomy, School of Physics and Astronomy, and Shanghai Key Laboratory for Particle Physics and Cosmology, Shanghai Jiao Tong University, Shanghai 200240, People's Republic of China}

\author[0000-0003-1132-8258]{Haojie Xu (许浩杰)}
\affiliation{Department of Astronomy, School of Physics and Astronomy, and Shanghai Key Laboratory for Particle Physics and Cosmology, Shanghai Jiao Tong University, Shanghai 200240, People's Republic of China}
\affiliation{Shanghai Astronomical Observatory, Chinese Academy of Sciences, Nandan Road 80, Shanghai 200240, People's Republic of China}

\author[0000-0002-0245-8547]{Yuyu Wang (王钰钰)}
\affiliation{Department of Astronomy, School of Physics and Astronomy, and Shanghai Key Laboratory for Particle Physics and Cosmology, Shanghai Jiao Tong University, Shanghai 200240, People's Republic of China}

\author{Antonios Katsianis}
\affiliation{Department of Astronomy, School of Physics and Astronomy, and Shanghai Key Laboratory for Particle Physics and Cosmology, Shanghai Jiao Tong University, Shanghai 200240, People's Republic of China}
\affiliation{School of Physics and Astronomy, Sun Yat-sen University, Zhuhai Campus, 2 Daxue Road, Xiangzhou District, Zhuhai, P. R. China}

\author[0009-0005-1980-8522]{Jiaxin Han (韩家信)}
\affiliation{Department of Astronomy, School of Physics and Astronomy, and Shanghai Key Laboratory for Particle Physics and Cosmology, Shanghai Jiao Tong University, Shanghai 200240, People's Republic of China}

\author{Min He (何敏)}
\affiliation{Department of Astronomy, School of Physics and Astronomy, and Shanghai Key Laboratory for Particle Physics and Cosmology, Shanghai Jiao Tong University, Shanghai 200240, People's Republic of China}

\author{Yunliang Zheng (郑云亮)}
\affiliation{Department of Astronomy, School of Physics and Astronomy, and Shanghai Key Laboratory for Particle Physics and Cosmology, Shanghai Jiao Tong University, Shanghai 200240, People's Republic of China}

\author{Qingyang Li (李清洋)}
\affiliation{Department of Astronomy, School of Physics and Astronomy, and Shanghai Key Laboratory for Particle Physics and Cosmology, Shanghai Jiao Tong University, Shanghai 200240, People's Republic of China}

\author{Yaru Wang (王雅茹)}
\affiliation{Department of Astronomy, School of Physics and Astronomy, and Shanghai Key Laboratory for Particle Physics and Cosmology, Shanghai Jiao Tong University, Shanghai 200240, People's Republic of China}

\author{Wensheng Hong (洪文生)}
\affiliation{Department of Astronomy, School of Physics and Astronomy, and Shanghai Key Laboratory for Particle Physics and Cosmology, Shanghai Jiao Tong University, Shanghai 200240, People's Republic of China}

\author{Jiaqi Wang (王佳琪)}
\affiliation{Department of Astronomy, School of Physics and Astronomy, and Shanghai Key Laboratory for Particle Physics and Cosmology, Shanghai Jiao Tong University, Shanghai 200240, People's Republic of China}

\author{Zhenlin Tan (谭镇林)}
\affiliation{Department of Astronomy, School of Physics and Astronomy, and Shanghai Key Laboratory for Particle Physics and Cosmology, Shanghai Jiao Tong University, Shanghai 200240, People's Republic of China}

\author[0000-0002-6684-3997]{Hu Zou (邹虎)}
\affiliation{National Astronomical Observatories, Chinese Academy of Sciences, A20 Datun Rd., Chaoyang District, Beijing, 100012, P.R. China}

\author{Johannes Ulf Lange}
\affiliation{Kavli Institute for Particle Astrophysics and Cosmology and Department of Physics, Stanford University, CA 94305, USA}
\affiliation{Department of Astronomy and Astrophysics, University of California, Santa Cruz, CA 95064, USA}
\affiliation{Department of Physics, University of Michigan, Ann Arbor, MI 48109, USA}
\affiliation{Leinweber Center for Theoretical Physics, University of Michigan, Ann Arbor, MI 48109, USA}

\author{ChangHoon Hahn}
\affiliation{Department of Astrophysical Sciences, Princeton University, Peyton Hall, Princeton, NJ 08544, USA}
\affiliation{Lawrence Berkeley National Laboratory, One Cyclotron Road, Berkeley, CA 94720, USA}

\author{Peter Behroozi}
\affiliation{Department of Astronomy and Steward Observatory, University of Arizona, Tucson, AZ 85721, USA}

\author{Jessica Nicole Aguilar}
\affiliation{Lawrence Berkeley National Laboratory, 1 Cyclotron Road, Berkeley, CA 94720, USA}
\author[0000-0001-6098-7247]{Steven Ahlen}
\affiliation{Physics Dept., Boston University, 590 Commonwealth Avenue, Boston, MA 02215, USA}
\author{David Brooks}
\affiliation{Department of Physics \& Astronomy, University College London, Gower Street, London, WC1E 6BT, UK}
\author{Todd Claybaugh}
\affiliation{Lawrence Berkeley National Laboratory, 1 Cyclotron Road, Berkeley, CA 94720, USA}
\author[0000-0002-5954-7903]{Shaun Cole}
\affiliation{Institute for Computational Cosmology, Department of Physics, Durham University, South Road, Durham DH1 3LE, UK}
\author[0000-0002-1769-1640]{Axel de la Macorra}
\affiliation{Instituto de F\'{\i}sica, Universidad Nacional Aut\'{o}noma de M\'{e}xico, Cd. de M\'{e}xico C.P. 04510, M\'{e}xico}
\author[0000-0002-5665-7912]{Biprateep Dey}
\affiliation{Department of Physics \& Astronomy and Pittsburgh Particle Physics, Astrophysics, and Cosmology Center (PITT PACC), University of Pittsburgh, 3941 O'Hara Street, Pittsburgh, PA 15260, USA}
\author{Peter Doel}
\affiliation{Department of Physics \& Astronomy, University College London, Gower Street, London, WC1E 6BT, UK}
\author[0000-0002-2890-3725]{Jaime E. Forero-Romero}
\affiliation{Departamento de F\'isica, Universidad de los Andes, Cra. 1 No. 18A-10, Edificio Ip, CP 111711, Bogot\'a, Colombia}
\affiliation{Observatorio Astron\'omico, Universidad de los Andes, Cra. 1 No. 18A-10, Edificio H, CP 111711 Bogot\'a, Colombia}
\author{Klaus Honscheid}
\affiliation{Department of Physics, The Ohio State University, 191 West Woodruff Avenue, Columbus, OH 43210, USA}
\affiliation{Center for Cosmology and AstroParticle Physics, The Ohio State University, 191 West Woodruff Avenue, Columbus, OH 43210, USA}
\author{Robert Kehoe}
\affiliation{Department of Physics, Southern Methodist University, 3215 Daniel Avenue, Dallas, TX 75275, USA}
\author[0000-0003-3510-7134]{Theodore Kisner}
\affiliation{Lawrence Berkeley National Laboratory, 1 Cyclotron Road, Berkeley, CA 94720, USA}
\author{Andrew Lambert}
\affiliation{Lawrence Berkeley National Laboratory, 1 Cyclotron Road, Berkeley, CA 94720, USA}
\author[0000-0003-4962-8934]{Marc Manera}
\affiliation{Institut de F\'{i}sica d’Altes Energies (IFAE), The Barcelona Institute of Science and Technology, Campus UAB, 08193 Bellaterra Barcelona, Spain}
\affiliation{Departament de F\'{i}sica, Serra H\'{u}nter, Universitat Aut\`{o}noma de Barcelona, 08193 Bellaterra (Barcelona), Spain}
\author[0000-0002-1125-7384]{Aaron Meisner}
\affiliation{NSF's NOIRLab, 950 N. Cherry Ave., Tucson, AZ 85719, USA}
\author{Ramon Miquel}
\affiliation{Institut de F\'{i}sica d’Altes Energies (IFAE), The Barcelona Institute of Science and Technology, Campus UAB, 08193 Bellaterra Barcelona, Spain}
\affiliation{Instituci\'{o} Catalana de Recerca i Estudis Avan\c{c}ats, Passeig de Llu\'{\i}s Companys, 23, 08010 Barcelona, Spain}
\author[0000-0002-2733-4559]{John Moustakas}
\affiliation{Department of Physics and Astronomy, Siena College, 515 Loudon Road, Loudonville, NY 12211, USA}
\author[0000-0001-6590-8122]{Jundan Nie}
\affiliation{National Astronomical Observatories, Chinese Academy of Sciences, A20 Datun Rd., Chaoyang District, Beijing, 100012, P.R. China}
\author{Claire Poppett}
\affiliation{Space Sciences Laboratory, University of California, Berkeley, 7 Gauss Way, Berkeley, CA 94720, USA}
\affiliation{Lawrence Berkeley National Laboratory, 1 Cyclotron Road, Berkeley, CA 94720, USA}
\author[0000-0001-5589-7116]{Mehdi Rezaie}
\affiliation{Department of Physics, Kansas State University, 116 Cardwell Hall, Manhattan, KS 66506, USA}
\author{Graziano Rossi}
\affiliation{Department of Physics and Astronomy, Sejong University, Seoul, 143-747, Korea}
\author[0000-0002-9646-8198]{Eusebio Sanchez}
\affiliation{CIEMAT, Avenida Complutense 40, E-28040 Madrid, Spain}
\author{Michael Schubnell}
\affiliation{Department of Physics, University of Michigan, Ann Arbor, MI 48109, USA}
\author[0000-0003-1704-0781]{Gregory Tarlé}
\affiliation{University of Michigan, Ann Arbor, MI 48109, USA}
\author{Benjamin Alan Weaver}
\affiliation{NSF's NOIRLab, 950 N. Cherry Ave., Tucson, AZ 85719, USA}
\author[0000-0002-4135-0977]{Zhimin Zhou}
\affiliation{National Astronomical Observatories, Chinese Academy of Sciences, A20 Datun Rd., Chaoyang District, Beijing, 100012, P.R. China}

\begin{abstract}
In this investigation, we leverage the combination of Dark Energy Spectroscopic Instrument Legacy imaging Surveys Data Release 9 (DESI LS DR9), Survey Validation 3 (SV3), and Year 1 (Y1) data sets to estimate the conditional luminosity and stellar mass functions (CLFs \& CSMFs) of galaxies across various halo mass bins and redshift ranges. To support our analysis, we utilize a realistic DESI Mock Galaxy Redshift Survey (MGRS) generated from a high-resolution Jiutian simulation. An extended halo-based group finder is applied to both MGRS catalogs and DESI observation.
By comparing the $r$ and $z$-band luminosity functions (LFs) and stellar mass functions (SMFs) derived using both photometric and spectroscopic data, we quantified the impact of photometric redshift (photo-z) errors on the galaxy LFs and SMFs, especially in the low redshift bin at low luminosity/mass end. By conducting prior evaluations of the group finder using MGRS, we successfully obtain a set of CLF and CSMF measurements from observational data. We find that at low redshift the faint end slopes of CLFs and CSMFs below $\sim 10^{9}\Lsunhh$ (or $\msunhh$) evince a compelling concordance with the subhalo mass functions. After correcting the cosmic variance effect of our local Universe following \citet{Chen2019}, the faint end slopes of the LFs/SMFs turn out to be also in good agreement with the slope of the halo mass function.
\end{abstract}



\section{Introduction}
\label{sec:Intro}
 
Over the past few decades, large galaxy surveys, such as the Two-degree Field Galaxy Redshift Survey (2dFGRS, \citealt{Colless1999}) and Sloan Digital Sky Survey (SDSS, \citealt{York2000}), have played a significant role in advancing our understanding of the galaxy formation and evolution. These surveys allow for various galaxy observable measurements, including the luminosity function (LF), stellar mass function (SMF), and two-point correlation function (2PCF) \citep{Norberg2002, blanton2003galaxy, zehavi2005luminosity, zehavi2011galaxy, wang2007cross, li2009distribution,zhao2000,Wang2021,2013ApJ...767...50M}.
Despite the absence of direct physical explanations of the galaxy formation and evolution, the statistical measurements provide essential constraints on multiple physical processes, including gravitational instability, gas cooling, star formation, merging, tidal stripping, heating, and feedback mechanisms. However, modeling the galaxy observables through physical processes remains a challenge, given the incomplete understanding of these processes \citep{Naab2017,Smercina2018,Katsianis2021,Sales2022}.

Under the hypothesis that galaxies form within dark matter halos, empirical halo models provide a straightforward way to model galaxy observables and infer the relationship between galaxies and their host halos \citep{Wechsler2018,Katsianis2023}. For instance, the halo occupation distribution (HOD, \citealt[][]{ jing1998spatial, peacock2000halo, tinker2005mass, Zheng2005, Zheng2009, brown2008red, 2015Zu, 2016Zu, 2018Zu, 2019Wang, 2020Alam, 2018Yuan, 2022Yuan}) infers the number of galaxies of a specific type in halos of different masses, and the conditional luminosity function (CLF, \citealt[][]{yang2003constraining, van2003linking, van2007towards,  yang2008galaxy, cacciato2009galaxy, Yang2012}) constrains the galaxy luminosity functions in halo of different masses. Additionally, the subhalo abundance matching (SHAM, \citealt[][]{Vale2004, conroy2006modeling, Behroozi2010, neistein2011tale, Guo2016}) links the number density of galaxies above a luminosity (or stellar mass) threshold to the number density of subhalos above a mass (or circular velocity) threshold. These empirical models have significantly enhanced our understandings of the galaxy formation and evolution processes. 

In addition to studies of the galaxy-halo connection through model fittings based on statistical measurements, an alternative method is introduced to identify individual dark matter halos observationally and measure the galaxy content within them. To this end, the halo-based group-finding algorithm \citep{yang2005halo} has particular advantages in grouping galaxies within the same dark matter halos and it has been extensively tested and applied to galaxy samples with spectroscopic redshifts \citep{yang2005halo, yang2007galaxy}. 
The original version of the group finder estimates halo mass using total luminosity inside each group. However, the accuracy may decrease for groups with few members, especially for a very shallow survey like 2MRS \citep{Huchra2012}. Other halo mass estimations are proposed to address this issue, for example, using the luminosity gap between the central and satellite or the luminosity-halo mass relation from hydrodynamic simulation \citep[]{Lu2015, Lim2017}. Other observables, such as the total luminosity of satellites in a halo, can also be used to estimate halo mass \citep[]{Tinker2020,Tinker2021}. Additionally, halo mass estimation can also be improved by considering the bimodality of star-forming or color \citep[]{Old2014, Old2015, Rodriguez-Puebla2015, Tinker2020}. These methodologies have demonstrated success in low-redshift surveys with high spectroscopic redshift completeness.

Apart from proposing different halo mass estimation methods, an alternative way to improve the halo mass estimation accuracy is to make use of faint galaxies in photometric redshift surveys. Within this framework, \citet{yang2021extended} extended the halo-based group finder so that it can deal with galaxies with photometric and spectroscopic redshifts simultaneously, which significantly broadened its application scope. Based on the increasing applicability of group finder, the CLFs and conditional stellar mass functions (CSMFs) have been successfully measured from the 2dFGRS, SDSS, HSC and DECaLS observations.  However, these measurements predominantly pertain to low redshifts and relatively luminous galaxies \citep[e.g.][]{yang2005galoccup, yang2008galaxy, yang2009galaxy, Lan2016, To2020, Meng2023, Golden-Marx2022, Wang2021, Thinker2021}.

In this study, we explore the LFs, SMFs, CLFs and CSMFs (central and satellite) of galaxies across different halo mass bins and redshift ranges through galaxy and group catalogues constructed by \citet{yang2021extended} from the DESI Image Legacy Surveys DR9 sample in redshift range of $z=[0, 1]$, which has a selection of apparent magnitudes down to $m_z=21$. We seek to evaluate the impact of photometric redshift and spectroscopic completness on the LFs and SMFs, with a particular focus on the faint end, by making full use of the first year of spectroscopic observation data.
Furthermore, we utilize a DESI mock galaxy redshift survey (MGRS, \citealt[]{Gu2024}) based on Jiutian, a high-precision N-body simulation (see more details in Section \ref{sec:data_Jiutian}), to perform the same statistical measurements using the same group finder employed in DESI observations, which facilitates the evaluation of systematic biases and allows for a more accurate investigation of the history of galaxy formation and evolution in future studies. Our galaxy samples are subject to observational effects, particularly in terms of the spectroscopic sampling. With the benefit from the extension version of the group finder, we expect for more reliable CLFs and CSMFs from DESI observations guided by simulated mock data.

In Section \ref{sec:Data}, we detail the photometric and spectroscopic data, including the construction of group catalogs and the sample selection for estimations of LFs, SMFs, CLFs, and CSMFs implemented in this paper. In Section \ref{sec:LF&SMF}, we investigate the impact of photo-z errors on the measurements of LFs and SMFs. In Section \ref{sec:Jiutian}, we use a MGRS to provide reliability verification of CLF measurements based on groups detected by the group finder. In Section \ref{sec:CLF&CSMF}, we present the CLFs and CSMFs based on the observational data. Finally, we discuss and summarize our results in Section \ref{sec:Discussion} and \ref{sec:Conclusion}.

Throughout this paper, we use $\Lambda$CDM cosmology with parameters that are consistent with the Planck 2018 results \citep{Planck2020}: $\Omega_{m}=0.315$, $\Omega_{\Lambda}=0.685$, $h=H_{0}/(100 km \ s^{-1}{\rm Mpc}^{-1}) = 0.674$ and $\sigma_{8} = 0.811$. Unless otherwise Specified, luminosity (stellar mass) and halo masses are presented in units of $\Lsunhh$($\msunhh$) and $\msunh$, respectively. The luminosity (stellar mass) functions are presented in units of $h^{3}{\rm Mpc}^{-3}d\log L$ ($h^{3}{\rm Mpc}^{-3}d\log M$), where $\log$ is the 10-based logarithm. The units of the conditional luminosity (stellar mass) function are $d\log L/$group and $d\log M/$group.


\section{Observational data}
\label{sec:Data}

\begin{figure*}[htb!]
\centering
\includegraphics[width=0.9\textwidth]{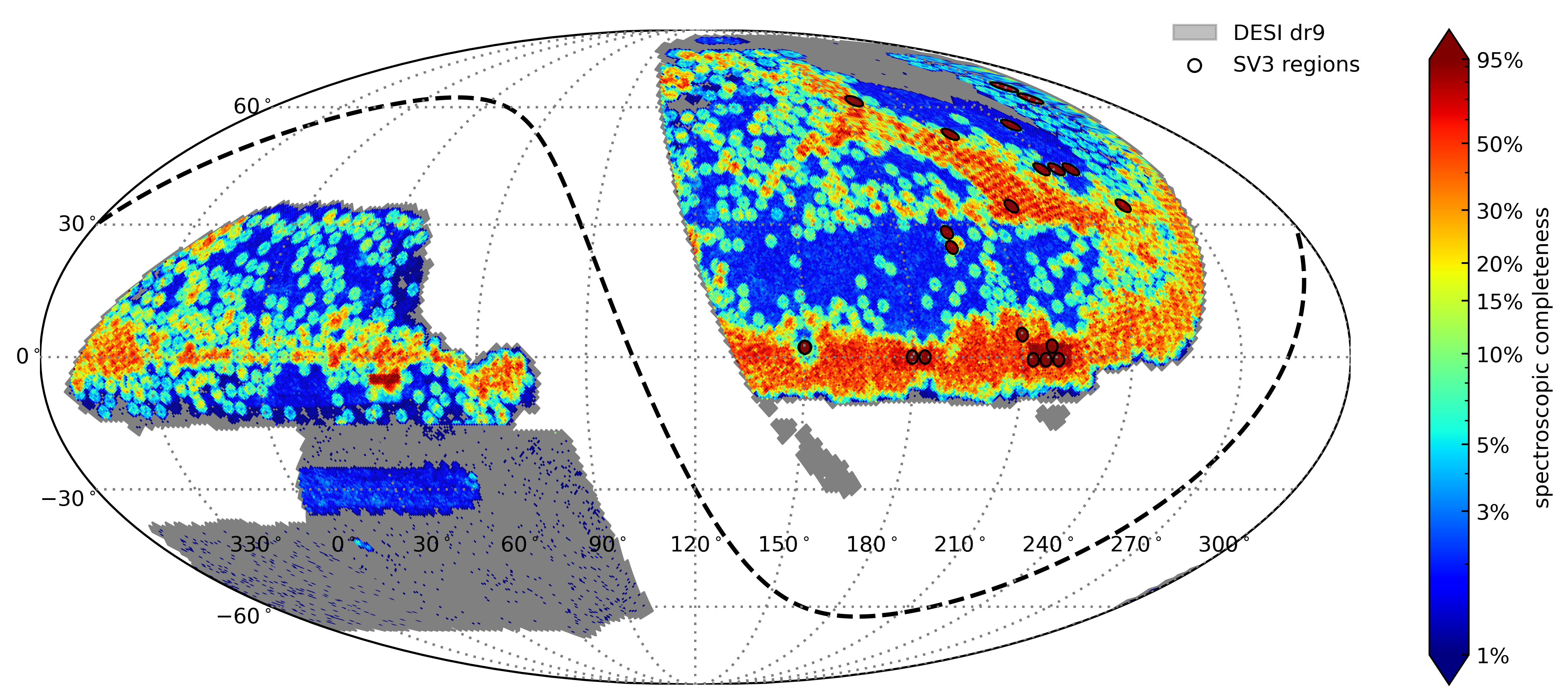} 
\caption{The sky coverage of the DESI spectroscopic data used in this study, where the spectroscopic completeness are calculated with respect to the total galaxies with $m_r\le 19.5$. The gray footprints are the ninth public data release of the DESI Legacy Imaging Surveys (DESI DR9), which we retains galactic latitude $|b| > 25$ degree. SV3 regions are marked by black circles, which exhibit extremely high completeness. Some discrete regions with low completeness are discretely distributed due to the collection of DESI Y1, SDSS, and other spectroscopic data. The back dashed line representing the galactic plane. }
\label{fig01} 
\end{figure*}

\begin{table*}
\caption{Sample definition and the related selection criteria.}
\begin{tabular}{lccccccc}
\hline
\hline
Sample ID      & sky coverage  & magnitude cut  & redshift cut & total  & central  & satellite  & spec-z percent \\ 
 & ($\rm deg^2$)& ($mag$) \\
\hline
SV3-r19.5   & 133    & $m_r\le 19.5$  & z$\le 0.6$ & 93943    & 73785    & 20158 (21.5\%)     & 99.2$\%$   \\
SV3-z19.0   & 133    & $m_z\le 19.0$  & z$\le 0.6$ & 120392   & 96016   & 24376 (20.2\%)     & 95.2$\%$   \\\hline
Y1-r19.5    & 12276  & $m_r\le 19.5$  & z$\le 0.6$ & 8464733  & 6484281  & 1980452 (23.4\%)   & 46.4$\%$   \\
Y1-z19.0    & 12276  & $m_z\le 19.0$  & z$\le 0.6$ & 10912062 & 8424354  & 2487708 (22.8\%)   & 44.3$\%$   \\\hline
\label{tab:sample}
\end{tabular}
\end{table*}

In this section, we describe the construction of galaxy samples from DESI observational data utilized in this study. The general overview and instrument of DESI can be found in a series of papers \citep{2016arXiv161100036D,2016arXiv161100037D,2022AJ....164..207A,2013arXiv1308.0847L,2023AJ....165....9S,2023arXiv230606310M}.
Overall, the galaxy sample is constructed by integrating a seed galaxy catalog, which is primarily based on the DESI Legacy imaging survey, with the data gathered from previous spectroscopic surveys and DESI up to the first year observation. Subsequently, group memberships are assigned using our extended adaptive halo-based group finder \citep[e.g.][]{yang2005halo, yang2007galaxy, yang2021extended}. 


\subsection{DESI Legacy Imaging Surveys DR9}
\label{sec:LS DR9}

 Legacy Surveys Data Release 9 (LS DR9) is the basis of the seed catalog, which includes three optical bands ({\it grz}) from the Beijing-Arizona Sky Survey (BASS, \citet{2017PASP..129f4101Z}), the Mayall Z-band Legacy Survey (MzLS) and the DECam Legacy Survey (DECaLS, \citet{Dey2019}). The DR9 also includes deeper optical data from the Dark Energy Survey (DES, \citet{2005astro.ph.10346T}). The optical bands of LS DR9 provide a 5$\sigma$ detection of 24/23.4/22.5 AB magnitude with a half-light radius of $0.45^{''}$. 
The target selections and survey validation of the DESI observational data are described in detail in \citet{2023arXiv230606307D,2023arXiv230606308D,2023ApJ...943...68L,2023AJ....165..124A,2022arXiv220808514C,2022arXiv220808512H,2023AJ....165...58Z,2023AJ....165..126R,2023ApJ...944..107C,2020RNAAS...4..188A,2020RNAAS...4..187R,2020RNAAS...4..181Z,2020RNAAS...4..180R,2020RNAAS...4..179Y,2023AJ....165...50M}.
We use the photometric redshift (photo-z) from The Photometric Redshifts for the Legacy Surveys (PRLS) catalogue \citep{Zhou2021}, who estimates photo-z by combining the optical and two mid-infrared photometry (W1 3.4$\mu$m and W2 4.6$\mu$m) from the {\it Wide-field Infrared Survey Explorer} (WISE). \citet{Zhou2021} demonstrates that the reliability of photoz estimation decreases beyond a $z$-band apparent magnitude of $m_z \simeq 21$. Therefore, the apparent magnitude of the $z$ band of our primary galaxy sample is limited to $m_z\le 21$. Although the overall selection of galaxies for this study closely follows that of \citet{yang2021extended}, a few modifications have been implemented to accommodate the transition from LS DR8 to LS DR9. Details about our sample selection are outlined below.

To mitigate the potential contamination, it is imperative to exclude stars and Active Galactic Nuclei (AGNs) from galaxy samples. Benefit from the morphological classification that identifies six distinct morphological types in DR9\footnote{This classification is facilitated by a software package called THE TRACTOR, as referenced in \citet{Lang2016}. THE TRACTOR is also used for source detection and optical photometry. See more details at \url{https://www.legacysurvey.org/dr9/description/##morphological-classification}}, we start by eliminating objects of the PSF and DUP \footnote{PSF stand for Point Spread Function, and DUP indicates Gaia sources that are coincident with an extended source.} types from the galaxy sample. The remianing extended sources with morphological classification of REX, EXP, DEV, and COMP\footnote{REX denotes round exponential galaxies with a variable radius. EXP indicates exponential profiles (spiral galaxies). DEV represents deVaucouleurs profiles (elliptical galaxies), and COMP indicates composite profiles combining deVaucouleurs and  exponential components.} consist our galaxy sample (similar to \citet{yang2021extended}).

To ensure photometric quality of our objects, constraints are imposed following the similar procedures as those described in \citet{Zou2019, Zhou2020, Zhou2021, Ruiz-Macias2020, Raichoor2020, Yeche2020, 2023ApJS..269....3M}. 
We require that each object has at least one exposure in each optical band. Objects located near the Galactic plane ($ b < 25^\circ$, where $b$ is the Galactic latitude) are eliminated to avoid regions of high stellar density.  
Additionally, the following bit numbers in the MASKBITS columns are used: 1 (close to Tycho-2 and GAIA bright stars), 5, 6, and 7 (close to objects which have the $\rm ALLMASK\_{[G,R,Z]}$ bits set), 8 (close to WISE W1 bright stars), 9 (close to WISE W2 bright stars), 11 (close to fainter GAIA stars), 12, and 13 (close to a local large galaxy and globular cluster, respectively). We use these selection conditions to remove the objects that are contaminated or blended.
In addition to the MASKBITS, other quality flags are employed to remove the flux contaminations from nearby sources (FRACFLUX) or masked pixels (FRACMASKED):
\begin{equation} \nonumber
{\rm FRACMASKED}_{\rm X} \ < \ 0.4
\end{equation} 
\begin{equation} \nonumber
{\rm FRACIN}_{\rm X} \ > \ 0.3
\end{equation}
\begin{equation} \nonumber
{\rm FRACFLAX}_{\rm X} \ < \ 0.5
\end{equation}
where ${\rm X} = g, r$, and $z$. The purpose of FRACIN is to select the objects for which a large fraction of the model flux is in the contiguous pixels where the model was fitted. Note that all the magnitudes used in this paper are in the AB system and have been corrected for Galactic extinction by using the Galactic transmission values provided in DR9. 

Following these criteria, we obtain a seed catalog of 138,315,649 galaxies. 
Most of these galaxies only contain photometric redshifts, which are the median values of the photo-z, z\_phot\_median, from the PRLS catalog. Approximately 3.7 million galaxies include spectroscopic redshifts collected from previous redshift surveys by \cite{Zhou2021} and \cite{Lim2017}. The redshifts and properties of galaxies are updated using the DESI spectroscopic data in Section \ref{sec:DESIupdate}.

\begin{table*}
\caption{ Mean  ${\rm K}$-corrections for $r$ and $z$ bands across various galaxy color categories (four color bins) over three redshift ranges. }
\begin{tabular}{llccccccccccc}
\hline
                       &       & \multicolumn{3}{c}{0.0-0.2} & &\multicolumn{3}{c}{0.2-0.4} & & \multicolumn{3}{c}{0.4-0.6} \\ \cline{3-5} \cline{7-9}  \cline{11-13}
                       &       & $a_{\mu}$   & $b_{\mu}$  & $c_{\mu}$ & & $a_{\mu}$    & $b_{\mu}$   & $c_{\mu}$  & & $a_{\mu}$    & $b_{\mu}$   & $c_{\mu}$    \\ \hline
\multirow{5}{*}{rband} & {color $\mu=1$ (blue)}  & -4.72   &  1.48    & -0.22  & & 1.06    & -0.49   & -0.24  & & -0.50    &  1.60   & -1.12   \\
                       &  {color $\mu=2$  }           & -1.52   &  0.98    & -0.19  & & 1.57    & -0.24   & -0.36   & &  3.41    & -1.30   & -0.64   \\
                       & {color $\mu=3$  }    &  0.40   &  0.78    & -0.19  & & 2.60    & -0.24   & -0.45   & &  4.29    & -1.28   & -0.88   \\
                       &   {color $\mu=4$  (red)}                      &  1.10   &  0.78    & -0.19  & & 2.27    &  0.22   & -0.55  & &  3.62    & -0.55   & -1.08   \\
                       & total                  & 0.13    &  0.80    & -0.19  & & 2.35    & -0.21   & -0.43 &  &  1.16    &  1.32   & -1.40   \\ \hline
\multirow{5}{*}{zband} & {color $\mu=1$  (blue)}  & 0.89    & -0.41  &  -0.07  & & 0.68    & -0.90   & -0.04  & & 2.60    & -2.00   & -0.08   \\
                       &  {color $\mu=2$  }        & 1.93    & -0.51   & -0.07 &  & 0.70    & -0.46   & -0.20  & & 2.65    & -2.00   & -0.10   \\
                       & {color $\mu=3$  }    & 1.97    & -0.33   & -0.09  & & 0.70    & -0.05   & -0.33  & & 1.72    & -1.11   & -0.31   \\
                       &     {color $\mu=4$  (red)}       & 1.35    & -0.07   & -0.11 &  & 0.21    &  0.42   & -0.42  & & 1.65    & -0.99   & -0.35   \\
                       & total                  & 2.24    & -0.45   & -0.08  & & 0.82    & -0.17   & -0.30  & & 1.82    & -1.19   & -0.29   \\ \hline
\label{tab:K-correction}
\end{tabular}
\end{table*}


\subsection{DESI spectroscopic data and group finder}
\label{sec:DESIupdate}

We make use of the most recent spectroscopic observation data (up to the first year observation) from the {\tt fastspecfit} Value-Added Catalogs (version 1.0\footnote{\url{https://fastspecfit.readthedocs.io/en/latest/fuji.html}\\ \url{https://fastspecfit.readthedocs.io/en/latest/guadalupe.html}\\ \url{https://fastspecfit.readthedocs.io/en/latest/iron.html}}), which contains three spectroscopic products, Fuji, Guadalupe, and Iron \citep[e.g.][]{Guy2023, Brodzeller2023, 2023ascl.soft08005M,2023arXiv230606309S}. 
Iron is the most comprehensive collection of spectral data available, containing 7.8 million galaxies spectra. We combine the fitting results of Fuji, Guadalupe, and Iron. For a galaxy with a unique target ID but multiple observations, we use the recommended ``best'' redshift with high quality in the combined catalogs across surveys and programs. The DESI Legacy Imaging Surveys (LS) guide the fiber assignment of the DESI spectroscopic survey. Therefore, our seed catalog, which primarily utilizes photometric redshift, can be updated with the measurements derived from DESI spectra. The updated seed catalog is used in the following analyses.

The extended group finder has been implemented to identify the membership of groups and estimate the group mass using both photometric and spectroscopic data. The group finder starts by considering each galaxy as a group candidate. The cumulative group luminosity functions can then be measured, and halo mass is assigned to each group using an abundance matching method. Subsequently, halo radius and line-of-sight velocity dispersion are estimated based on this halo mass. Beginning with the most massive group, the member galaxies are identified in the region where the galaxy number density contrasts is higher than a specific threshold. With the updated membership of groups, both the group center and the total luminosity of the group can be updated, and the algorithm goes back to the step of measuring the group luminosity function and assigning halo mass. The iteration continues until the mass-to-light ratios have converged.

We have noticed that 7.8\% galaxies lacks measurements in one or more of the five bands of $g/r/z/W1/W2$, despite having assigned photo-z values. 
Taken into account the much larger photo-z uncertainty of these galaxies than the typical values at $\sigma_{photo}\sim (0.01+0.015z)(1+z)$, we have excluded them from the group finding process and used a weight to correct this factor in the LF/SMF and CLF/CSMF measurements (see Section \ref{sec:LF} for more details).

By applying the extended halo-based group finder to the updated seed catalog, a group catalog is created that covers a wide range of redshifts and halo masses. According to \citet{yang2021extended}, the extended halo-based group finder is highly successful in identifying more than 60 percent of the members in almost 90 percent of halos with masses greater than $10^{12.5}\msunh$ for galaxies with magnitudes $m_z\le21$ and photometric redshifts in the range $0 < {\rm z} \le 1.0$ in the DESI legacy imaging surveys\footnote{Better performance can be achieved in a spectroscopic redshift sample \citep[see][]{yang2007galaxy}.}. The group catalog provides useful information about the host halo properties of galaxies, such as the halo mass and local environments, and the central/satellite classification. Since most galaxies only have photometric redshifts, the entire sample is classified as LS DR9.


\subsection{Sample selection}
\label{sec:samplesel}

In this paper, we consider three regions in the sky coverage with different levels of spectroscopic completeness: 
\begin{itemize}
    \item The whole region of the updated seed galaxy catalog with a magnitude limit $m_z\le 21$, constructed from LS DR9 and the DESI spectroscopic survey, which is primarily used to find the groups;   
    \item The Y1 region with a wide sky coverage of 12276 $deg^{2}$ and an approximate $\sim 45\%$ spectral completeness after the Bright Galaxy Survey (BGS) selection ($m_r\le 19.5$ and $m_z\le 19$), which is used to measure the CLFs and CSMFs.
    \item The SV3 region with the highest spectral completeness (over 95\%) after the bright magnitude cuts above, which is used as a benchmark to investigate the effect of photo-z on the LF and SMF measurements.  
\end{itemize}

The SV3 region and the Y1 region are defined by a set of rosettes and tiles using caps at a given radius, respectively, while the overall footprint of the seed catalog is mapped using the healpix tool \citep{2005ApJ...622..759G}. This tool divides the spherical surface into subdivisions, each of which covers the same surface area. We set the parameter $\rm nside = 256$, which corresponds to $\rm 5.246\times 10^{-2} deg^2$ per subdivision. Subdivisions with at least one galaxy are treated as part of the footprint. Figure \ref{fig01} shows the sky coverage of the spectroscopic data used in this study. The SV3 region is marked by black circles, while the Y1 region is indicated by the color gradient area from cyan to red. The color coding represents the spectroscopic completeness calculated with respect to the total galaxies of $m_r\le 19.5$. The remaining regions, characterized by gray and blue, primarily lack spectroscopic data and contribute mainly to the group finder.

To ensure a higher rate of spectroscopic redshift completeness and a reasonable sampling, we restrict our analysis to the galaxy sample with $m_r\le 19.5$ and $z\le 0.6$, which is in line with the selection criteria of the DESI BGS survey. It has been verified that galaxies in our sample with $m_r\le 19.5$ are almost identical to the targets of the BGS Bright sample conducted by \citet{Hahn2023, 2024Hahn}. Taking into account the magnitude difference between the $r$ and $z$ band, this roughly corresponds to $m_z\le 19.0$. Combining the footprint and apparent magnitude cut, we define four sub-samples for our analysis and list the details in Table \ref{tab:sample}. 
In the first two,  Y1-r19.5 and Y1-z19.0, nearly 45\% of the galaxies have spectroscopic redshifts. For simplicity, we use Y1-BGS to refer to Y1-r19.5 or Y1-z19.0 depending on the band we are using.

Within the Y1 region, the DESI 1\% Survey (also known as SV3, which includes three spectroscopic productions but covers a smaller area) has a significantly higher spectroscopic completeness, which is marked by black circles in Figure \ref{fig01}, and the spectroscopic completeness of the remaining area is relatively lower.
The SV3 strategy was mainly focused on guiding and validating the survey design. To achieve a high level of completeness, additional passes were performed for each of the 20 rosettes within SV3. These passes covered an area of more than $\rm 7\ deg^2$, extending up to 1.45 degrees from the center of each discrete region. Among these 20 rosettes, one rosette with celestial coordinates (194.75, 28.20) is centered in the Coma cluster. As we will demonstrate in Figure \ref{fig:19circle_LF} in Appendix \ref{appendix:A}, the inclusion of this rosette will significantly enhance the LFs (or SMFs) at $L\sim 10^8\Lsunhh$, resulting in considerable cosmic variances. Consequently, we opted to exclude this rosette from our LF and SMF measurements. 
By applying the BGS selections mentioned above, the spectroscopic redshift completeness is overall larger than 95\% with $\rm 133\ deg^2$ sky coverage. Here again, for simplicity, we use SV3-BGS to represent the third and fourth sub-samples SV3-z19.0 or SV3-r19.5, depending on the band we are using. Despite the relatively limited sky coverage and the disconnection of SV3-BGS, the high redshift completeness allows for the verification of the impact of photo-z errors on LF and SMF measurements.

The group catalogs of the SV3 and Y1 subsamples are extracted from those of LS DR9. The detailed selection criteria and the total number of galaxies, as well as the number of central and satellite galaxies in our four subsamples, are listed in Table~\ref{tab:sample}.  This enables us to distinguish the contribution of central and satellites to the conditional luminosity function, affording valuable insights into the distribution of galaxies within groups. 


\subsection{Galaxy luminosity and stellar mass }
\label{sec:lum_mem}

Following \citet{yang2021extended}, for each galaxy passed to the group finder, we use the following function to convert
apparent magnitude to absolute magnitude according to its redshift.
\begin{equation} \label{eq:mag_eachgal}
{\rm M^j}_{\rm X} - 5\log h = 
{\rm m}_{\rm X}  - {\rm DM}(z_{obs}) - {\rm K}^{j}_{\rm X}(z_{obs})\,
\end{equation}
where $\rm X$ stands for the particular band ($r$ or $z$) we adopted. ${\rm K}^{j}_{\rm X}$ represents each galaxy's ${\rm K}$-correction to $\rm X$-band shifted by the band-shift redshift, j, where $j = 0.1, 0.3$, or 0.5 obtained from the `Kcorrect' model (eg. v4\_3) described in \citet{Blanton2007}. $DM(z_{obs})$ is the distance module corresponding to the redshift $z_{obs}$ defined as
\begin{equation} \label{eq:distmeas}
{\rm DM}({\rm z_{obs}}) = 5 \log D_L({\rm z_{obs}}) + 25,
\end{equation}
with ${\rm D_L}(z_{obs})$ being the luminosity distance in unit of $\mpch$.  
The luminosity of each galaxy is then calculated using the following formula: 
\begin{equation} \label{eq:lum}
\log_{10} L^{j}_{\rm X} = 0.4*({\rm M^{j}}_{\odot}-{\rm M^{j}}_{\rm X})\,
\end{equation}
For a better consideration of the absolute magnitude of the sun after K-correction, ${\rm M^{j}_\odot}$, we use the fitting results of K-correction in narrow redshift bins from $z = 0$ to maximum redshift. These K-correction values at typical redshifts are listed below, which are consistent with ${\rm M^{j=0}_\odot}$ being 4.61 and 4.5 in the $r$ and $z$ bands, respectively \citep{Willmer2018}.
\begin{equation} \nonumber
{\rm K}^{[0.1,0.3,0.5]}_{\odot, r}(0.0)=[-0.19,-0.42,-0.75]
\end{equation}
\begin{equation} \nonumber
{\rm K}^{[0.1,0.3,0.5]}_{\odot, z}(0.0)=[-0.08,-0.26,-0.33]
\end{equation}
 
Apart from the ${\rm K}$ corrections, \citet{blanton2007k} also provide an estimation of the stellar mass for each galaxy with fast spectral analysis and stellar composition estimation.
They combine heterogeneous data (including broad-band fluxes at various redshifts) in order to determine the properties of the subspace of galaxy spectra.  They restrict the space of possible spectra to those predicted from the high-resolution stellar population synthesis model of \citet{Bruzual2003} and the nebular emission line models of \citet{Kewley2001}. This approach yields a natural theoretical interpretation of the results in terms of star formation histories.
The consistency of this method with the mass-to-light ratio method proposed by \citet{kauffmann2003stellar} has been checked and verified. However, due to the limitations of redshift uncertainty and the templates used, the stellar masses obtained by the ${\rm K}$-correction code should be handled with caution. The stellar masses may be overestimated to some extent due to the long star formation history assumed, particularly at high redshifts. Nevertheless, it still provides a quick and efficient estimation of the stellar mass. 
Furthermore, we have also adopted the SED code
CIGALE \citep{Boquien2020}, which is an alternative method for estimating stellar mass and is applied to DESI observations recently in \citet{Xu2022}. The stellar masses derived using CIGALE align with our results, maintaining consistency within a margin of error of 0.1 dex across various stellar mass scales.


\section{The impact of photo-z errors on the LF and SMF measurements}
\label{sec:LF&SMF}

Prior to examining the CLF and CSMF, this section will analyze the impact of photo-z error on the measurements of LFs and SMFs using the Y1-BGS and SV3-BGS sub-samples.

\begin{figure*}[!htp]
\centering
\includegraphics[width=0.9\textwidth]{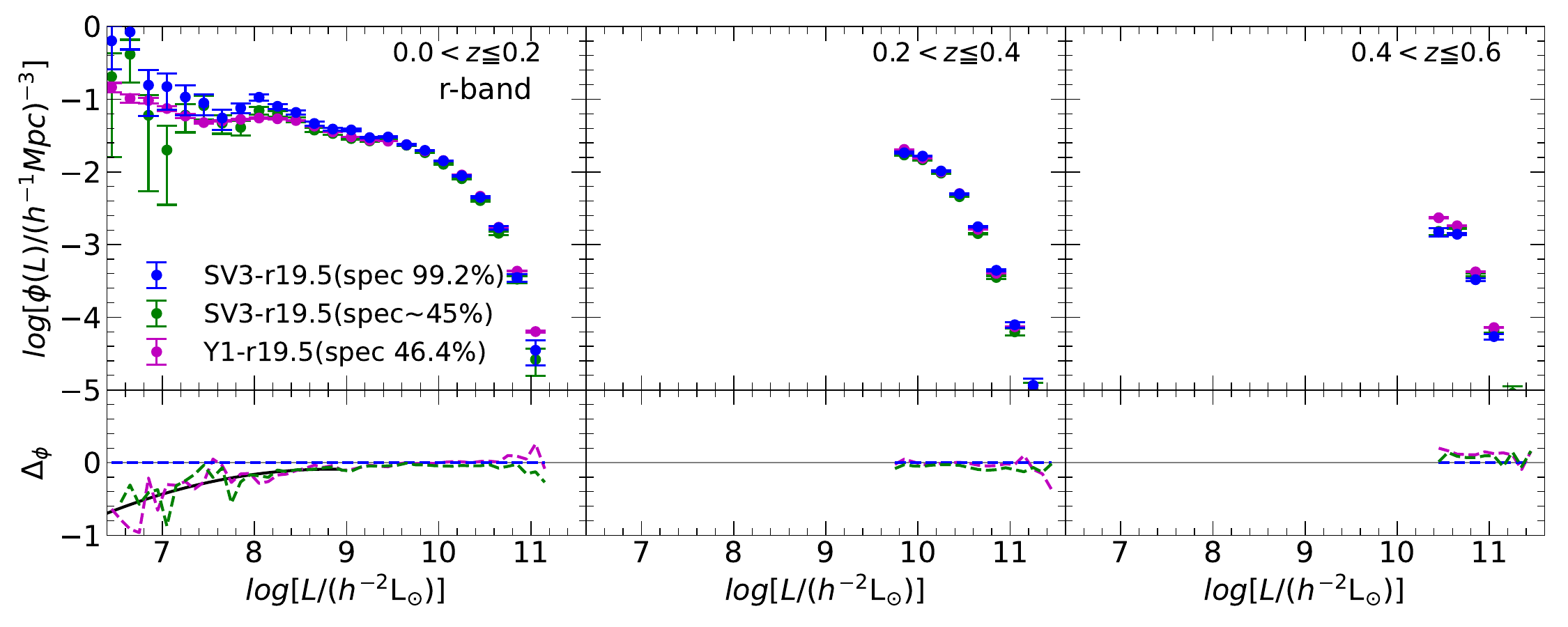}
\includegraphics[width=0.9\textwidth]{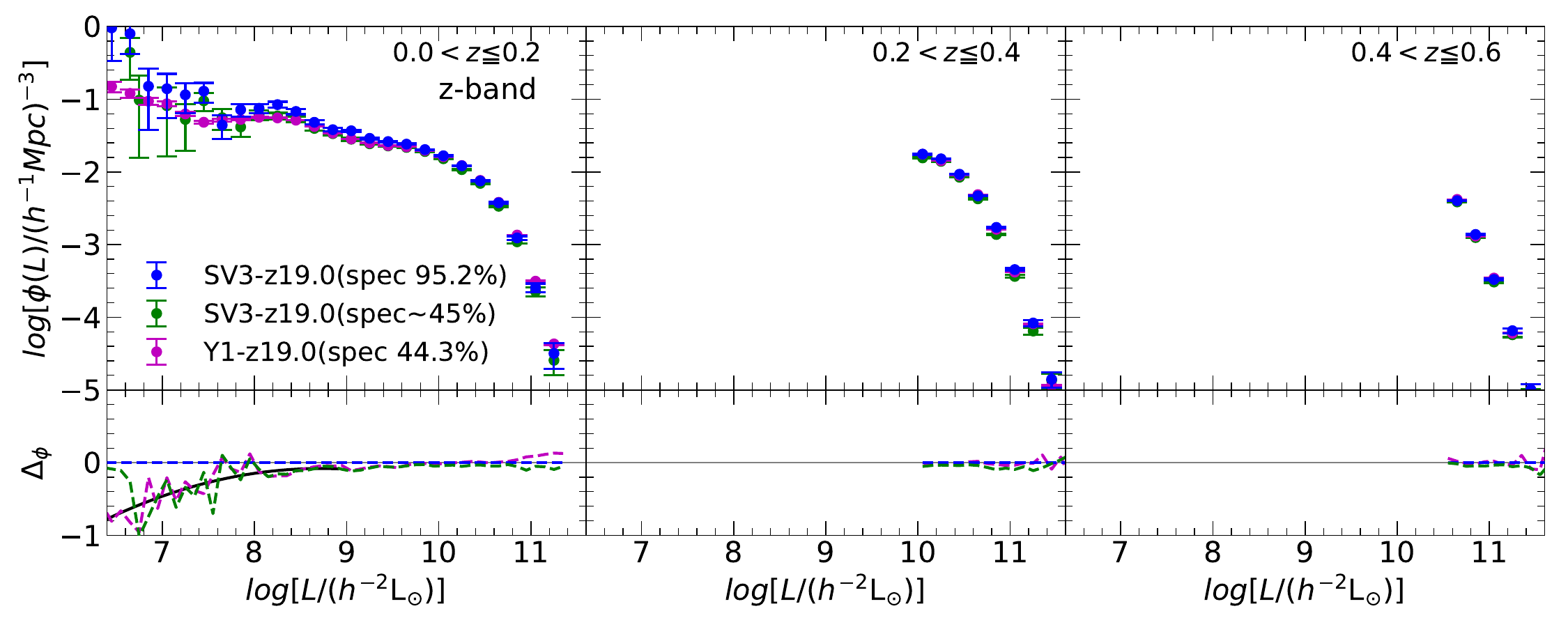}
\includegraphics[width=0.9\textwidth]{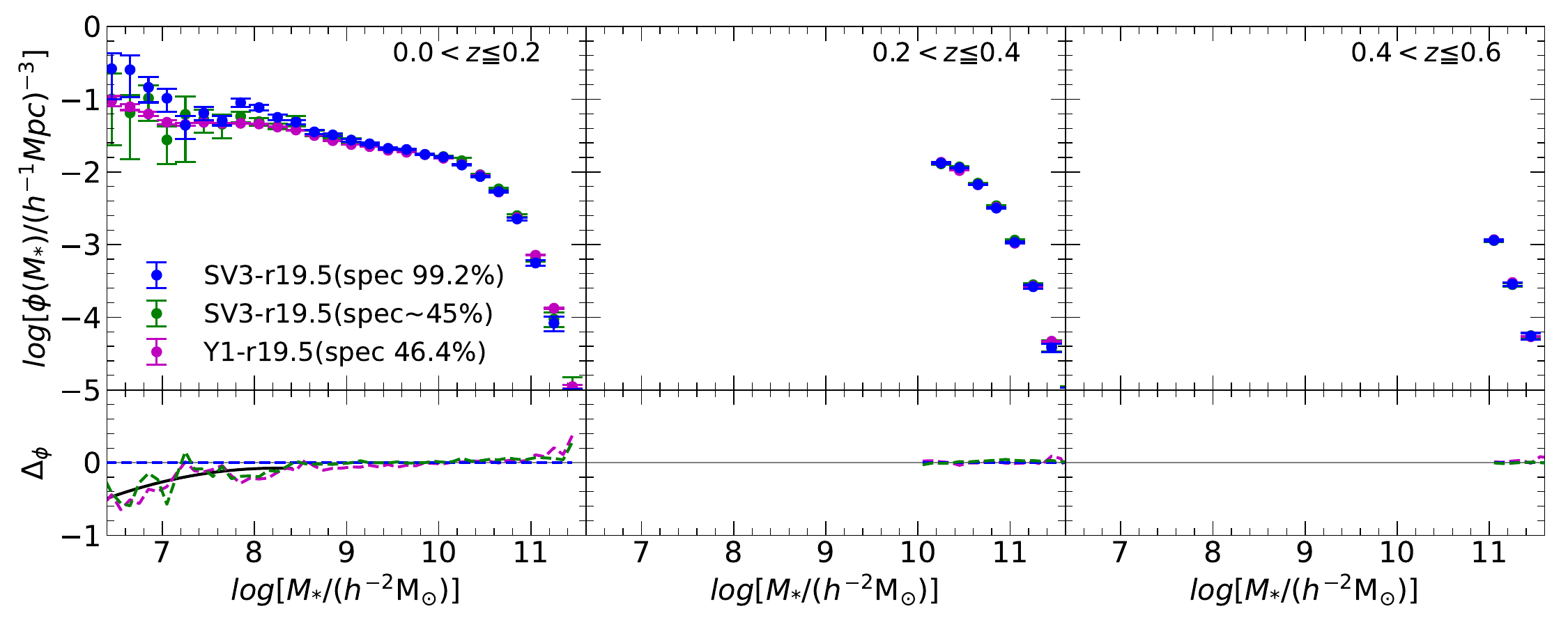}
\caption{Galaxy LFs and SMFs in three redshift bins as indicated for two observational sub-samples after ${\rm K}$-correction. The upper, middle, and bottom row panels show the results for $r$-band,  $z$-band luminosity, and stellar mass, respectively. The blue and magenta dots in each panel are for SV3-r19.5 (SV3-z19.0) and Y1-r19.5 (Y1-z19.0) sub-samples, respectively. The green ones are the results of a degraded SV3-BGS(spec $\sim45\%$) sub-sample. The error bars are obtained from the standard deviation of 200 bootstrap resamplings. The lower portions of panels show deviations between SV3-BGS and other samples. The black solid lines are fitting formulas of the suppression factors which are $-0.103(\log{L})^{2}+1.815(\log{L})-8.100$, $-0.130(\log{L})^{2}+2.258(\log{L})-9.925$, $-0.123(\log{M_{*}})^{2}+2.027(\log{M_{*}})-8.425$, for the upper, middle and lower panels, respectively.  } \label{dr9plots}
\end{figure*}


\subsection{Galaxy luminosity functions}
\label{sec:LF}

We employ a standard $V_{\rm max}$ approach to calculate the LFs for $r$-band and $z$-band. We start by removing any galaxies with apparent magnitudes beyond the magnitude limit from a magnitude-limited galaxy sample ($m_r\le 19.5$ and $m_z\le 19$). We then select the redshift range $[{\rm z}_1, {\rm z}_2]$ within which the galaxies are used for the LF measurements. For a galaxy with a specific absolute magnitude, we determine the maximum redshift ${\rm z}_{\rm max}$ below which the galaxy can be observed by adopting an approximate ${\rm K}$-correction in this process. Finally, we calculate the effective volume for the galaxy according to the redshift range $[{\rm z}_1, {\rm min}({\rm z}_2,{\rm z}_{\rm max})]$. 

In general, the approximate ${\rm K}$-correction applied to calculate
$z_{\rm max}$ can be described as

\begin{equation} \label{eq:mag_zmax}
{\rm M^j}_{\rm X} - 5\log h = 
{\rm m}^{\rm limit}_{\rm X}  - {\rm DM}(z_{max}) - {\rm K'}^{j}_{\rm X}({\rm z_{max}})\,,
\end{equation}
where ${\rm m}^{\rm limit}_{\rm X}$ is the magnitude limit, 19.5 for the $r$-band and 19 for the $z$-band, and 
\begin{equation} \label{eq:Kcorr_mean1}
{\rm K'}^{j}_{\rm X}({\rm z_{max}}) = {\rm K}^{j}_{\rm X}({\rm z_{obs}}) + {\rm \bar{K}}^{j}_{\rm X}({\rm z_{max}}) - {\rm \bar{K}}^{j}_{\rm X}({\rm z_{obs}})
\end{equation}
is an approximate K-correction when relocated a galaxy at $z_{\rm obs}$ to $z_{\rm max}$, which can be estimated by the mean K-correction as function of galaxy color and redshift at these two redshifts
\begin{equation} \label{eq:Kcorr_mean2}
{\rm \bar{K}}^{j}_{\rm X}({\rm z}) = \Sigma \ \omega_{\mu}(r-z) (a_{\mu}{\rm z}^2+b_{\mu}{\rm z}+c_{\mu})\,.
\end{equation}
Following Rodriguez-Puebla et al. (2020), galaxies within each redshift bin are initially categorized into four $r-z$ color bins indicating by $\mu$, and the weight $\omega_{\mu}$ is determined with the distance of a given galaxy to the bin center on the color-magnitude plane. The redshift-dependent coefficients $a_\mu$, $b_\mu$, and $c_\mu$ for the mean K-correction in each color bin $\mu$ are derived by fitting the K-correction from \citet{blanton2007k}, and the results are shown in Table \ref{tab:K-correction}.
The K correction, ${\rm K'}^{j}_{\rm X}({\rm z_{max}})$, is then applied to determine the corresponding $z_{max}$.

To ensure the quality of observation data, we calculated the fraction $f_{\rm comp} (m_z)$, which represents the ratio of the number of galaxies with observations at five wavelengths to the number of all galaxies, using healpix \citep{Gorski2005} as a function of the apparent magnitude of the $z$ band and the color of the galaxy in each pixel. This factor is then used to adjust for the selection incompleteness in the subsequent analysis.

We first measure the LFs of the Y1-BGS (Y1-r19.5 or Y1-z19.0) and SV3-BGS (SV3-r19.5 or SV3-z19.0)
sub-samples in three redshift bins. The upper and middle panels of Figure \ref{dr9plots} show the results for the $r$- and $z$-bands, respectively.
The results of the $r$-band and the $z$-band are shown in the upper and middle panels of Figure \ref{dr9plots}, respectively. 
The LFs of the two bands are generally consistent, although the $r$-band luminosity is slightly lower than the $z$-band luminosity. 
The blue dots with error bars represent the SV3-BGS sub-sample, which has a small sky coverage (133 $\rm deg^2$) and most of the galaxies with spectroscopic redshifts, thus providing the most reliable LF measurements. The magenta dots show the results of the Y1-BGS sub-sample, which has a large sky coverage (12276 $\rm deg^2$) and has roughly half of its galaxies with spectroscopic redshifts. Overall, the LFs of Y1-BGS and SV3-BGS show very good agreement with each other with negligible differences in all redshift bins. The main deviations from the SV3-BGS sub-sample only become prominent as the luminosity decreases below $L \le 10^{8.5}\Lsunhh$. 

To investigate the influence of photo-z errors on LF measurements, we created a degraded SV3-BGS sample (spectra percentage of galaxies $\sim45\%$, after here spec $\sim45\%$) which closely matches the spectral completeness of Y1-BGS by randomly replace 55\% spectroscopic redshifts of the galaxies with their original photometric redshifts before obtaining the spectroscopic observation of SV3-BGS and show the results with green dots. The error bars are estimated using the bootstrap method, which randomly resamples the original galaxy sample with replacement while keeping the total number of galaxies unchanged.
To demonstrate the deviations more clearly, the logarithmic differences between the LFs of Y1-BGS (spec $\sim45\%$) and SV3-BGS are shown in the lower part of each panel in Figure \ref{dr9plots}. The degraded SV3-BGS and Y1-BGS sub-samples show good agreement within 1-$\sigma$, demonstrating that the differences between the SV3-BGS and Y1-BGS results are indeed caused by the photoz error.

We use a quadratic function (represented by the solid black curves) to fit the deviation of LFs from the original to degraded SV3-BGS sub-samples, which will be used to correct the LF deviations observed in the Y1-BGS sub-samples. The fitted deviations for the $r$-band and $z$-band are $-0.103(\log{L})^{2}+1.815(\log{L})-8.100$ and $-0.130(\log{L})^{2}+2.258(\log{L})-9.925$, respectively. To avoid overcrowding of LF and SMF data points, only half of the luminosity bins listed in Table \ref{LF_SMF_result} are shown. For $\Delta \phi$, all luminosity bins are shown with dashed lines.

\begin{table*}[!ht]
\centering
\caption{Values of the  galaxy LFs and SMFs obtained from SV3-r19.5 and SV3-z19.0 sub-samples in different redshift bins.}
\resizebox{18cm}{!}{
\begin{tabular}{cccccccccc}
\hline
\toprule
\multirow{2}{*}{\begin{tabular}[c]{@{}c@{}}$L$ or $M_{*}$ \\ 
$[\Lsunhh]$ ~~ $[\msunhh]$\end{tabular}} & \multicolumn{3}{c}{\begin{tabular}[c]{@{}c@{}}$\log\Phi(L)-rband$ \\ 
$[h^{3}{\rm Mpc}^{-3}d\log L]$ \end{tabular}} & \multicolumn{3}{c}{\begin{tabular}[c]{@{}c@{}}$\log\Phi(L)-zband$ \\ 
$[h^{3}{\rm Mpc}^{-3}d\log L]$ \end{tabular}} & \multicolumn{3}{c}{\begin{tabular}[c]{@{}c@{}}$\log\Phi(M_{*})$ \\ 
$[h^{3}{\rm Mpc}^{-3}d\log M_{*}]$ \end{tabular}} \\ \cline{2-10} 
& 0.0-0.2 & 0.2-0.4 & 0.4-0.6  & 0.0-0.2 & 0.2-0.4 & 0.4-0.6 & 0.0-0.2 & 0.2-0.4 & 0.4-0.6      \\ \hline 
6.45 & $-0.1967_{-0.3927}^{+0.2028}$ & $-_{ }^{ }$ & $-_{ }^{ }$  & $-0.0186_{-0.4561}^{+0.2175}$ & $-_{ }^{ }$ & $-_{ }^{ }$ & $-0.5806_{-0.4174}^{+0.2088}$ & $-_{ }^{ }$ & $-_{ }^{ }$  \\
6.55 & $-0.0175_{-0.2001}^{+0.1364}$ & $-_{ }^{ }$ & $-_{ }^{ }$  & $-0.1533_{-0.4216}^{+0.2098}$ & $-_{ }^{ }$ & $-_{ }^{ }$ & $-0.3140_{-0.3421}^{+0.1889}$ & $-_{ }^{ }$ & $-_{ }^{ }$  \\
6.65 & $-0.0737_{-0.2408}^{+0.1540}$ & $-_{ }^{ }$ & $-_{ }^{ }$  & $-0.0981_{-0.2782}^{+0.1682}$ & $-_{ }^{ }$ & $-_{ }^{ }$ & $-0.5941_{-0.3756}^{+0.1983}$ & $-_{ }^{ }$ & $-_{ }^{ }$  \\
6.75 & $-0.0670_{-0.1597}^{+0.1165}$ & $-_{ }^{ }$ & $-_{ }^{ }$  & $0.00546_{-0.2111}^{+0.1414}$ & $-_{ }^{ }$ & $-_{ }^{ }$ & $-0.5580_{-0.3123}^{+0.1798}$ & $-_{ }^{ }$ & $-_{ }^{ }$  \\
6.85 & $-0.8062_{-0.4245}^{+0.2105}$ & $-_{ }^{ }$ & $-_{ }^{ }$  & $-0.8216_{-0.6036}^{+0.2432}$ & $-_{ }^{ }$ & $-_{ }^{ }$ & $-0.8368_{-0.2065}^{+0.1394}$ & $-_{ }^{ }$ & $-_{ }^{ }$  \\
6.95 & $-0.4230_{-0.1866}^{+0.1301}$ & $-_{ }^{ }$ & $-_{ }^{ }$  & $-0.3453_{-0.1930}^{+0.1331}$ & $-_{ }^{ }$ & $-_{ }^{ }$ & $-0.8681_{-0.2465}^{+0.1563}$ & $-_{ }^{ }$ & $-_{ }^{ }$  \\
7.05 & $-0.8264_{-0.3174}^{+0.1814}$ & $-_{ }^{ }$ & $-_{ }^{ }$  & $-0.8552_{-0.3999}^{+0.2046}$ & $-_{ }^{ }$ & $-_{ }^{ }$ & $-0.9860_{-0.1861}^{+0.1298}$ & $-_{ }^{ }$ & $-_{ }^{ }$  \\
7.15 & $-0.8980_{-0.2335}^{+0.1510}$ & $-_{ }^{ }$ & $-_{ }^{ }$  & $-0.6113_{-0.2360}^{+0.1520}$ & $-_{ }^{ }$ & $-_{ }^{ }$ & $-1.1888_{-0.1864}^{+0.1300}$ & $-_{ }^{ }$ & $-_{ }^{ }$  \\
7.25 & $-0.9709_{-0.2530}^{+0.1588}$ & $-_{ }^{ }$ & $-_{ }^{ }$  & $-0.9390_{-0.2546}^{+0.1594}$ & $-_{ }^{ }$ & $-_{ }^{ }$ & $-1.3580_{-0.1868}^{+0.1302}$ & $-_{ }^{ }$ & $-_{ }^{ }$  \\
7.35 & $-0.9116_{-0.1908}^{+0.1321}$ & $-_{ }^{ }$ & $-_{ }^{ }$  & $-0.8751_{-0.1931}^{+0.1332}$ & $-_{ }^{ }$ & $-_{ }^{ }$ & $-1.2177_{-0.1659}^{+0.1197}$ & $-_{ }^{ }$ & $-_{ }^{ }$  \\
7.45 & $-1.0523_{-0.1650}^{+0.1193}$ & $-_{ }^{ }$ & $-_{ }^{ }$  & $-0.8890_{-0.1554}^{+0.1142}$ & $-_{ }^{ }$ & $-_{ }^{ }$ & $-1.1906_{-0.1003}^{+0.0814}$ & $-_{ }^{ }$ & $-_{ }^{ }$  \\
7.55 & $-1.3414_{-0.2053}^{+0.1388}$ & $-_{ }^{ }$ & $-_{ }^{ }$  & $-1.1349_{-0.1966}^{+0.1348}$ & $-_{ }^{ }$ & $-_{ }^{ }$ & $-1.2387_{-0.1021}^{+0.0826}$ & $-_{ }^{ }$ & $-_{ }^{ }$  \\
7.65 & $-1.2584_{-0.1613}^{+0.1173}$ & $-_{ }^{ }$ & $-_{ }^{ }$  & $-1.3557_{-0.1902}^{+0.1318}$ & $-_{ }^{ }$ & $-_{ }^{ }$ & $-1.2919_{-0.0680}^{+0.0588}$ & $-_{ }^{ }$ & $-_{ }^{ }$  \\
7.75 & $-1.0426_{-0.0716}^{+0.0614}$ & $-_{ }^{ }$ & $-_{ }^{ }$  & $-1.2047_{-0.1309}^{+0.1004}$ & $-_{ }^{ }$ & $-_{ }^{ }$ & $-1.1429_{-0.0629}^{+0.0549}$ & $-_{ }^{ }$ & $-_{ }^{ }$  \\
7.85 & $-1.1208_{-0.0707}^{+0.0608}$ & $-_{ }^{ }$ & $-_{ }^{ }$  & $-1.1441_{-0.0990}^{+0.0806}$ & $-_{ }^{ }$ & $-_{ }^{ }$ & $-1.0476_{-0.0602}^{+0.0529}$ & $-_{ }^{ }$ & $-_{ }^{ }$  \\
7.95 & $-1.1091_{-0.0643}^{+0.0560}$ & $-_{ }^{ }$ & $-_{ }^{ }$  & $-1.3577_{-0.1073}^{+0.0859}$ & $-_{ }^{ }$ & $-_{ }^{ }$ & $-1.1211_{-0.0511}^{+0.0457}$ & $-_{ }^{ }$ & $-_{ }^{ }$  \\
8.05 & $-0.9730_{-0.0425}^{+0.0387}$ & $-_{ }^{ }$ & $-_{ }^{ }$  & $-1.1244_{-0.0700}^{+0.0603}$ & $-_{ }^{ }$ & $-_{ }^{ }$ & $-1.1136_{-0.0416}^{+0.0379}$ & $-_{ }^{ }$ & $-_{ }^{ }$  \\
8.15 & $-1.0043_{-0.0443}^{+0.0402}$ & $-_{ }^{ }$ & $-_{ }^{ }$  & $-1.0548_{-0.0551}^{+0.0489}$ & $-_{ }^{ }$ & $-_{ }^{ }$ & $-1.1513_{-0.0362}^{+0.0334}$ & $-_{ }^{ }$ & $-_{ }^{ }$  \\
8.25 & $-1.0976_{-0.0370}^{+0.0341}$ & $-_{ }^{ }$ & $-_{ }^{ }$  & $-1.0740_{-0.0438}^{+0.0398}$ & $-_{ }^{ }$ & $-_{ }^{ }$ & $-1.2495_{-0.0393}^{+0.0360}$ & $-_{ }^{ }$ & $-_{ }^{ }$  \\
8.35 & $-1.1191_{-0.0335}^{+0.0311}$ & $-_{ }^{ }$ & $-_{ }^{ }$  & $-1.0818_{-0.0378}^{+0.0348}$ & $-_{ }^{ }$ & $-_{ }^{ }$ & $-1.3290_{-0.0362}^{+0.0334}$ & $-_{ }^{ }$ & $-_{ }^{ }$  \\
8.45 & $-1.1817_{-0.0289}^{+0.0271}$ & $-_{ }^{ }$ & $-_{ }^{ }$  & $-1.1681_{-0.0389}^{+0.0357}$ & $-_{ }^{ }$ & $-_{ }^{ }$ & $-1.3157_{-0.0330}^{+0.0307}$ & $-_{ }^{ }$ & $-_{ }^{ }$  \\
8.55 & $-1.2463_{-0.0262}^{+0.0247}$ & $-_{ }^{ }$ & $-_{ }^{ }$  & $-1.2526_{-0.0323}^{+0.0301}$ & $-_{ }^{ }$ & $-_{ }^{ }$ & $-1.4163_{-0.0314}^{+0.0293}$ & $-_{ }^{ }$ & $-_{ }^{ }$  \\
8.65 & $-1.3337_{-0.0281}^{+0.0264}$ & $-_{ }^{ }$ & $-_{ }^{ }$  & $-1.3174_{-0.0293}^{+0.0275}$ & $-_{ }^{ }$ & $-_{ }^{ }$ & $-1.4494_{-0.0281}^{+0.0264}$ & $-_{ }^{ }$ & $-_{ }^{ }$  \\
8.75 & $-1.3414_{-0.0227}^{+0.0215}$ & $-_{ }^{ }$ & $-_{ }^{ }$  & $-1.3747_{-0.0278}^{+0.0261}$ & $-_{ }^{ }$ & $-_{ }^{ }$ & $-1.4216_{-0.0255}^{+0.0241}$ & $-_{ }^{ }$ & $-_{ }^{ }$  \\
8.85 & $-1.4107_{-0.0205}^{+0.0195}$ & $-_{ }^{ }$ & $-_{ }^{ }$  & $-1.4189_{-0.0250}^{+0.0236}$ & $-_{ }^{ }$ & $-_{ }^{ }$ & $-1.4911_{-0.0204}^{+0.0195}$ & $-_{ }^{ }$ & $-_{ }^{ }$  \\
8.95 & $-1.3755_{-0.0135}^{+0.0131}$ & $-_{ }^{ }$ & $-_{ }^{ }$  & $-1.4706_{-0.0217}^{+0.0207}$ & $-_{ }^{ }$ & $-_{ }^{ }$ & $-1.5238_{-0.0175}^{+0.0168}$ & $-_{ }^{ }$ & $-_{ }^{ }$  \\
9.05 & $-1.4224_{-0.0147}^{+0.0142}$ & $-_{ }^{ }$ & $-_{ }^{ }$  & $-1.4340_{-0.0143}^{+0.0139}$ & $-_{ }^{ }$ & $-_{ }^{ }$ & $-1.5626_{-0.0199}^{+0.0190}$ & $-_{ }^{ }$ & $-_{ }^{ }$  \\
9.15 & $-1.4838_{-0.0134}^{+0.0130}$ & $-_{ }^{ }$ & $-_{ }^{ }$  & $-1.4839_{-0.0156}^{+0.0151}$ & $-_{ }^{ }$ & $-_{ }^{ }$ & $-1.5694_{-0.0163}^{+0.0157}$ & $-_{ }^{ }$ & $-_{ }^{ }$  \\
9.25 & $-1.5282_{-0.0116}^{+0.0113}$ & $-_{ }^{ }$ & $-_{ }^{ }$  & $-1.5382_{-0.0141}^{+0.0137}$ & $-_{ }^{ }$ & $-_{ }^{ }$ & $-1.6140_{-0.0150}^{+0.0145}$ & $-_{ }^{ }$ & $-_{ }^{ }$  \\
9.35 & $-1.5252_{-0.0092}^{+0.0090}$ & $-_{ }^{ }$ & $-_{ }^{ }$  & $-1.5743_{-0.0125}^{+0.0122}$ & $-_{ }^{ }$ & $-_{ }^{ }$ & $-1.6170_{-0.0133}^{+0.0129}$ & $-_{ }^{ }$ & $-_{ }^{ }$  \\
9.45 & $-1.5195_{-0.0096}^{+0.0094}$ & $-_{ }^{ }$ & $-_{ }^{ }$  & $-1.5824_{-0.0107}^{+0.0104}$ & $-_{ }^{ }$ & $-_{ }^{ }$ & $-1.6761_{-0.0122}^{+0.0119}$ & $-_{ }^{ }$ & $-_{ }^{ }$  \\
9.55 & $-1.5771_{-0.0078}^{+0.0076}$ & $-_{ }^{ }$ & $-_{ }^{ }$  & $-1.5641_{-0.0090}^{+0.0088}$ & $-_{ }^{ }$ & $-_{ }^{ }$ & $-1.6597_{-0.0109}^{+0.0106}$ & $-_{ }^{ }$ & $-_{ }^{ }$  \\
9.65 & $-1.6260_{-0.0076}^{+0.0075}$ & $-_{ }^{ }$ & $-_{ }^{ }$  & $-1.6177_{-0.0083}^{+0.0082}$ & $-_{ }^{ }$ & $-_{ }^{ }$ & $-1.6916_{-0.0093}^{+0.0091}$ & $-_{ }^{ }$ & $-_{ }^{ }$  \\
9.75 & $-1.6443_{-0.0069}^{+0.0068}$ & $-_{ }^{ }$ & $-_{ }^{ }$  & $-1.6675_{-0.0072}^{+0.0071}$ & $-_{ }^{ }$ & $-_{ }^{ }$ & $-1.7008_{-0.0090}^{+0.0088}$ & $-_{ }^{ }$ & $-_{ }^{ }$  \\
9.85 & $-1.7039_{-0.0063}^{+0.0062}$ & $-1.7373_{-0.0150}^{+0.0145}$ & $-_{ }^{ }$  & $-1.6927_{-0.0058}^{+0.0057}$ & $-_{ }^{ }$ & $-_{ }^{ }$ & $-1.7598_{-0.0082}^{+0.0081}$ & $-_{ }^{ }$ & $-_{ }^{ }$  \\
9.95 & $-1.7649_{-0.0065}^{+0.0064}$ & $-1.7571_{-0.0076}^{+0.0074}$ & $-_{ }^{ }$  & $-1.7187_{-0.0063}^{+0.0062}$ & $-1.7403_{-0.0153}^{+0.0148}$ & $-_{ }^{ }$ & $-1.7736_{-0.0094}^{+0.0092}$ & $-_{ }^{ }$ & $-_{ }^{ }$  \\
10.05 & $-1.8477_{-0.0077}^{+0.0076}$ & $-1.7825_{-0.0062}^{+0.0061}$ & $-_{ }^{ }$  & $-1.7778_{-0.0071}^{+0.0070}$ & $-1.7549_{-0.0088}^{+0.0086}$ & $-_{ }^{ }$ & $-1.7950_{-0.0061}^{+0.0060}$ & $-_{ }^{ }$ & $-_{ }^{ }$  \\
10.15 & $-1.9458_{-0.0085}^{+0.0083}$ & $-1.8634_{-0.0054}^{+0.0054}$ & $-_{ }^{ }$  & $-1.8437_{-0.0066}^{+0.0065}$ & $-1.7721_{-0.0061}^{+0.0060}$ & $-_{ }^{ }$ & $-1.8548_{-0.0076}^{+0.0075}$ & $-1.8597_{-0.0155}^{+0.0150}$ & $-_{ }^{ }$  \\
10.25 & $-2.0531_{-0.0106}^{+0.0104}$ & $-1.9878_{-0.0053}^{+0.0052}$ & $-_{ }^{ }$  & $-1.9164_{-0.0074}^{+0.0073}$ & $-1.8201_{-0.0049}^{+0.0048}$ & $-_{ }^{ }$ & $-1.9053_{-0.0084}^{+0.0082}$ & $-1.8782_{-0.0102}^{+0.0100}$ & $-_{ }^{ }$  \\
10.35 & $-2.1708_{-0.0120}^{+0.0117}$ & $-2.1360_{-0.0056}^{+0.0056}$ & $-_{ }^{ }$  & $-2.0191_{-0.0096}^{+0.0093}$ & $-1.9171_{-0.0046}^{+0.0045}$ & $-_{ }^{ }$ & $-1.9751_{-0.0100}^{+0.0098}$ & $-1.9156_{-0.0075}^{+0.0074}$ & $-_{ }^{ }$  \\
10.45 & $-2.3505_{-0.0145}^{+0.0140}$ & $-2.3025_{-0.0060}^{+0.0059}$ & $-_{ }^{ }$  & $-2.1249_{-0.0107}^{+0.0105}$ & $-2.0319_{-0.0045}^{+0.0045}$ & $-_{ }^{ }$ & $-2.0641_{-0.0102}^{+0.0100}$ & $-1.9441_{-0.0063}^{+0.0062}$ & $-_{ }^{ }$  \\
10.55 & $-2.5483_{-0.0180}^{+0.0173}$ & $-2.5129_{-0.0069}^{+0.0068}$ & $-2.6235_{-0.0169}^{+0.0163}$  & $-2.2473_{-0.0124}^{+0.0121}$ & $-2.1759_{-0.0043}^{+0.0043}$ & $-2.2477_{-0.0189}^{+0.0181}$ & $-2.1691_{-0.0118}^{+0.0115}$ & $-2.0596_{-0.0059}^{+0.0059}$ & $-_{ }^{ }$  \\
10.65 & $-2.7684_{-0.0239}^{+0.0226}$ & $-2.7542_{-0.0089}^{+0.0087}$ & $-2.8579_{-0.0133}^{+0.0129}$  & $-2.4249_{-0.0142}^{+0.0137}$ & $-2.3303_{-0.0052}^{+0.0051}$ & $-2.3964_{-0.0076}^{+0.0074}$ & $-2.2714_{-0.0125}^{+0.0122}$ & $-2.1738_{-0.0067}^{+0.0066}$ & $-_{ }^{ }$  \\
10.75 & $-3.1338_{-0.0387}^{+0.0356}$ & $-3.0326_{-0.0119}^{+0.0115}$ & $-3.1627_{-0.0144}^{+0.0139}$  & $-2.6385_{-0.0209}^{+0.0199}$ & $-2.5196_{-0.0065}^{+0.0064}$ & $-2.6203_{-0.0072}^{+0.0071}$ & $-2.4683_{-0.0172}^{+0.0166}$ & $-2.3222_{-0.0059}^{+0.0058}$ & $-_{ }^{ }$  \\
10.85 & $-3.4533_{-0.0563}^{+0.0498}$ & $-3.3540_{-0.0183}^{+0.0176}$ & $-3.4830_{-0.0222}^{+0.0211}$  & $-2.9081_{-0.0250}^{+0.0236}$ & $-2.7634_{-0.0114}^{+0.0111}$ & $-2.8605_{-0.0075}^{+0.0073}$ & $-2.6473_{-0.0221}^{+0.0210}$ & $-2.4985_{-0.0069}^{+0.0068}$ & $-_{ }^{ }$  \\
10.95 & $-3.8053_{-0.0846}^{+0.0708}$ & $-3.7260_{-0.0279}^{+0.0262}$ & $-3.8880_{-0.0273}^{+0.0257}$  & $-3.2384_{-0.0397}^{+0.0364}$ & $-3.0322_{-0.0123}^{+0.0120}$ & $-3.1485_{-0.0096}^{+0.0093}$ & $-2.8964_{-0.0263}^{+0.0248}$ & $-2.7192_{-0.0089}^{+0.0087}$ & $-2.7514_{-0.0350}^{+0.0324}$  \\
11.05 & $-4.4542_{-0.2073}^{+0.1397}$ & $-4.1033_{-0.0398}^{+0.0364}$ & $-4.2659_{-0.0388}^{+0.0356}$  & $-3.5929_{-0.0657}^{+0.0570}$ & $-3.3436_{-0.0192}^{+0.0184}$ & $-3.4799_{-0.0132}^{+0.0128}$ & $-3.2529_{-0.0427}^{+0.0388}$ & $-2.9692_{-0.0110}^{+0.0107}$ & $-2.9404_{-0.0152}^{+0.0147}$  \\
11.15 & $-4.6842_{-0.2565}^{+0.1602}$ & $-4.6675_{-0.0962}^{+0.0787}$ & $-4.7231_{-0.0653}^{+0.0567}$  & $-4.0093_{-0.1153}^{+0.0910}$ & $-3.7102_{-0.0253}^{+0.0239}$ & $-3.7984_{-0.0226}^{+0.0215}$ & $-3.5683_{-0.0579}^{+0.0510}$ & $-3.2492_{-0.0153}^{+0.0148}$ & $-3.2121_{-0.0247}^{+0.0234}$  \\
11.25 & $-_{ }^{ }$ & $-4.9351_{-0.1156}^{+0.0912}$ & $-5.1914_{-0.1146}^{+0.0906}$  & $-4.4976_{-0.2117}^{+0.1417}$ & $-4.0792_{-0.0443}^{+0.0402}$ & $-4.1849_{-0.0326}^{+0.0303}$ & $-4.0808_{-0.1091}^{+0.0871}$ & $-3.5797_{-0.0256}^{+0.0242}$ & $-3.5444_{-0.0223}^{+0.0212}$  \\
11.35 & $-_{ }^{ }$ & $-5.4608_{-0.2963}^{+0.1745}$ & $-5.4673_{-0.1798}^{+0.1267}$  & $-5.0903_{-0.6093}^{+0.2440}$ & $-4.6180_{-0.0803}^{+0.0677}$ & $-4.7458_{-0.0657}^{+0.0571}$ & $-4.4571_{-0.2180}^{+0.1444}$ & $-3.9017_{-0.0407}^{+0.0372}$ & $-3.9025_{-0.0272}^{+0.0256}$  \\
11.45 & $-_{ }^{ }$ & $-5.8772_{-0.5012}^{+0.2265}$ & $-6.1826_{-0.7476}^{+0.2603}$  & $-_{ }^{ }$ & $-4.8526_{-0.1177}^{+0.0925}$ & $-4.9955_{-0.0902}^{+0.0747}$ & $-5.3309_{-0.6926}^{+0.3430}$ & $-4.4195_{-0.0627}^{+0.0548}$ & $-4.2583_{-0.0462}^{+0.0417}$  \\
11.55 & $-_{ }^{ }$ & $-_{ }^{ }$ & $-_{ }^{ }$  & $-_{ }^{ }$ & $-5.5604_{-0.3428}^{+0.1891}$ & $-5.4857_{-0.1686}^{+0.1211}$ & $-_{ }^{ }$ & $-4.8287_{-0.0967}^{+0.0790}$ & $-4.7906_{-0.0651}^{+0.0566}$  \\
11.65 & $-_{ }^{ }$ & $-_{ }^{ }$ & $-6.4058_{-1.1864}^{+0.3149}$  & $-_{ }^{ }$ & $-5.8315_{-0.7039}^{+0.2558}$ & $-6.5010_{-1.4491}^{+0.3086}$ & $-_{ }^{ }$ & $-5.0822_{-0.1480}^{+0.1102}$ & $-5.2429_{-0.1397}^{+0.1055}$  \\
11.75 & $-_{ }^{ }$ & $-_{ }^{ }$ & $-_{ }^{ }$  & $-_{ }^{ }$ & $-_{ }^{ }$ & $-6.4879_{-1.2941}^{+0.2898}$ & $-_{ }^{ }$ & $-5.5104_{-0.3261}^{+0.1841}$ & $-5.8058_{-0.2709}^{+0.1655}$  \\  \\ \hline
\end{tabular}}
\label{LF_SMF_result}
\end{table*}


\subsection{Galaxy stellar mass functions}
\label{sec:SMF}

The stellar mass of galaxies is one of the most important properties in the study of galaxy evolution and cosmic structure evolution. It can be reliably measured using SED modeling \citep{Conroy2009,Conroy2013,Song2023},  and is more commonly used in theoretical studies. In measuring the galaxy SMFs, we use the $r$-band apparent and absolute magnitudes to calculate z$_{\rm max}$ for each galaxy in our SV3-r19.5 and Y1-r19.5 sub-samples. The resulting SMFs are shown in the lower panels of Figure \ref{dr9plots}. As we have tested, the  SMFs of SV3-BGS and Y1-BGS using the CIGALE and K-correction code are in good agreement with each other within 1-$\sigma$ error bars.

These SMFs display similar trends to the LFs, especially the enhancement starting around $M_{\ast}\sim 10^{8.5} \msunhh$. 
Such kind of enhancement in SMFs at the low mass end was also reported in a recent study carried out by \citet{Gao2023}.  Finally, similar to the LFs, we also fit the deviation of SMFs of degraded SV3-BGS (spec $\sim45\%$) sub-sample with respect to the initial SV3-BGS sub-sample using a quadratic function form, denoted by the solid black lines in Figure \ref{dr9plots}. This fitting result, $-0.123(\log{M_{*}})^{2}+2.027(\log{M_{*}})-8.425$, can be applied to correct the SMF suppression of Y1-BGS at low mass end. The results of the luminosity functions and stellar mass functions obtained from SV3-BGS are provided in Table~\ref{LF_SMF_result} for reference.

\begin{figure*}[!ht]
\centering
\includegraphics[width=0.9\textwidth]{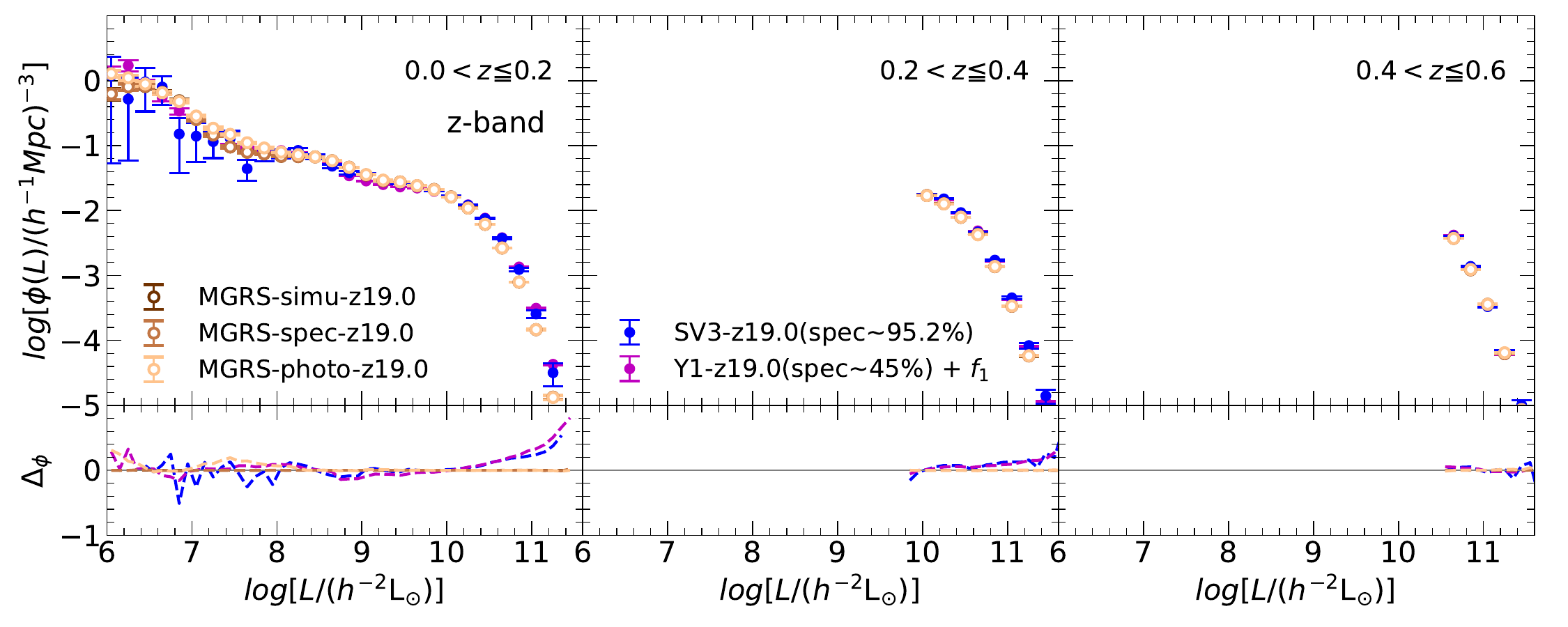}
\caption{LFs ($z$-band) of DESI LS DR9 MGRS. Here MGRS-simu (dark brown), MGRS-spec (light brown), and MGRS-photo (orange) represent mock galaxies in real space, with spectroscopic redshifts, and mixed spectroscopic and photometric redshifts, respectively. Also, we show LFs of observation (SV3-z19.0 and Y1-z19.0) for comparison. MGRS-simu and MGRS-spec are virtually indistinguishable. The error bars are obtained from the standard deviation of 200 bootstrap resamplings. The lower small panels show deviations between MGRS-simu-z19.0 and other samples. \label{LFJiutian}}
\end{figure*}


\section{Testing the reliability of CLF measurements using MGRS}
\label{sec:Jiutian}

Due to various selection effects, observational data sometimes might lead to biased measurements.
To assess the influence of the group finder and photo-z error on the CLF and CSMF measurements, we create an MGRS from the Jiutian simulation with the same sky coverage as the LS DR9. The luminosity of the MGRS is adjusted to match the $z$-band LFs from the Y1-z19.0 sample after a correction factor is applied to the SV3-z19.0 sample. Two sets of mock group catalogs are then generated from the MGRS using the extended halo-based group finder with either spectroscopic or photometric redshifts.

\begin{figure*}[!htp]
\centering
\includegraphics[width=1\textwidth]{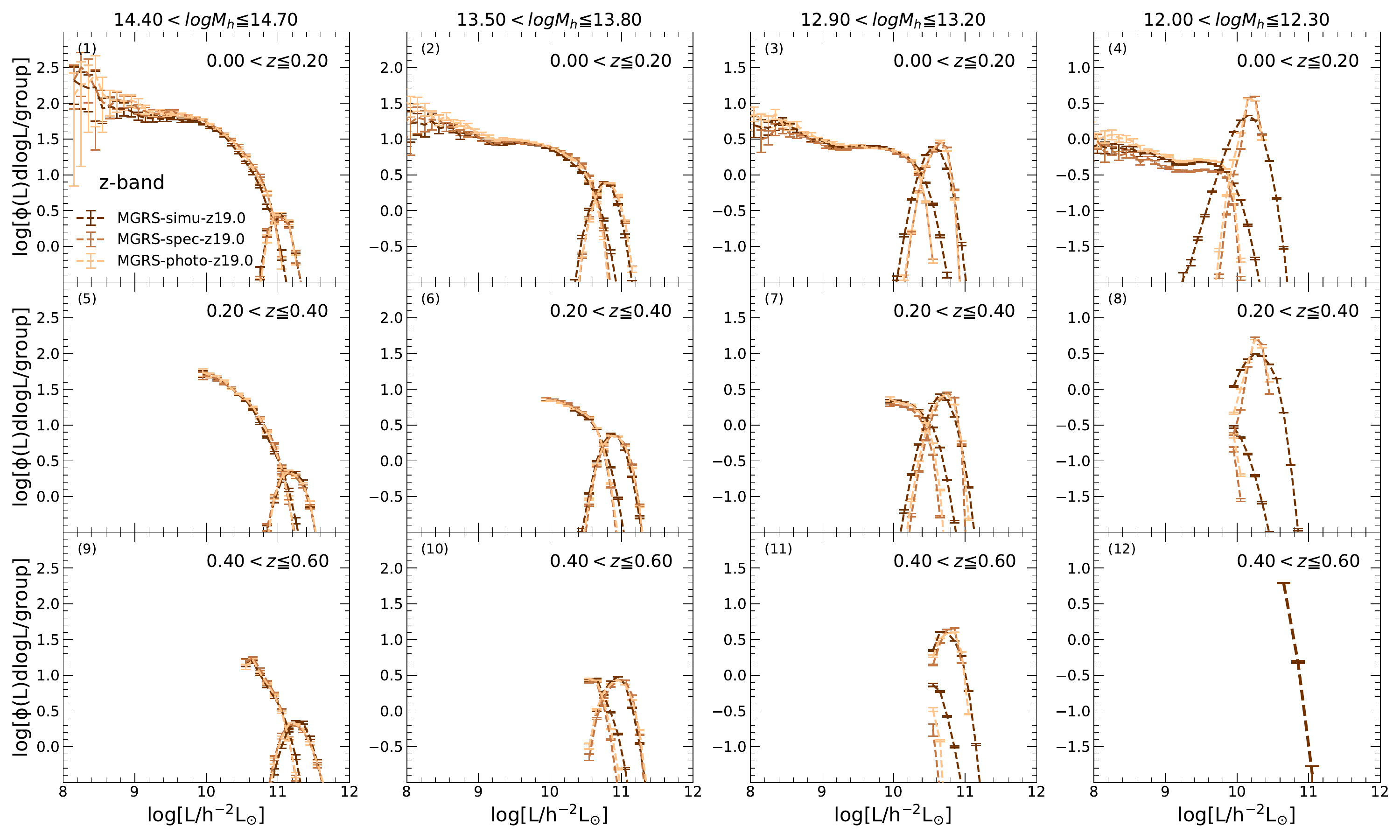}
\caption{CLFs ($z$-band) obtained from the MGRS. MGRS-simu, MGRS-spec, and MGRS-photo represent the true CLFs, the ones are obtained from mock groups with spectroscopic and mixed spectroscopic and photometric redshifts, respectively. Each column corresponds to a different halo mass bin and each row stands for a different redshift bin. The errors are obtained from the standard deviation of 200 bootstrap resamplings.  \label{CLFJiutian}}
\end{figure*}


\subsection{Constructing MGRS from Jiutian simulation}
\label{sec:data_Jiutian}

We employed a high-resolution dark-matter-only $N$-body simulation from the Jiutian simulation suite, specifically designed for the optical surveys conducted by the Chinese Space-station Survey Telescope (CSST, \citealt{Zhan2011,Zhan2018}), to construct our MGRS  \citep{Gu2024}. The three main runs of the simulation suite are based on Planck-2018 cosmology \citep{Planck18}, with parameters listed in Section~\ref{sec:Intro}, evolving $6144^3$ particles in boxes of 0.3, 1, and 2 $\mathrm{Gpc}/h$ on a side. Extension runs spanning various cosmologies and constrained runs reproducing the large scale structure in observation will also be available. The simulation we adopted is the main run of   $1\,\mathrm{Gpc}/h$ with particle mass of $m_p=3.723\times10^8 \msunh$ using the \textsc{Gadget}-3 code \citep{Gadget, Gadget2}. The simulation began at an initial redshift of $127$, producing 128 snapshots to $z=0$.  The Friends-of-Friends algorithm~\citep{Davis1985} with a linking length of $0.2$ times the mean inter-particle separation was used to identify dark matter halos. The \textsc{HBT+} code~\citep{han2012resolving,han2018hbt+}\footnote{\url{https://github.com/Kambrian/HBTplus}} was then used to identify subhalos and their evolution histories.

An observer is placed at a reference location in the box, and each snapshot is replicated periodically to create a subhalo light cone. The orbit of each subhalo is then interpolated over time to determine the time and location at which the subhalo should be observed. The properties of the subhalo are then interpolated to the intersecting time to generate a lightcone catalog of subhalos.  method allows for the precise recovery of the mass function and clustering of (sub)halos across different redshifts. Furthermore, the large size of our simulation box minimizes the duplication of subhalos, particularly at lower redshifts. Using the Jiutian subhalo light cone catalog, we assign a $z$-band galaxy luminosity to each subhalo with a subhalo abundance matching method SHAM, \citealt[][]{Vale2004,Reddick2013} that incorporates a luminosity scatter $\sigma_{\log L}=0.15$dex at fixed subhalo mass. The subhalo mass we adopted is the maximum mass along the accretion history. The cumulative LFs used for the abundance matching are directly measured from the Y1-z19.0 sample with narrow redshift bins. To assign luminosity to each subhalo at a particular redshift, we interpolated the LFs at different redshifts. It is worth noting that $V_{\rm peak}$ may be more preferred than the peak mass as subhalo mass indicator for SHAM in the literature \citep{Reddick2013,Lehmann2017,Dragomir2018}, and the comparison of the two will be discussed in more detail in a forthcoming paper (Xu et al. 2024 in preparation).

Our MGRS for LS DR9 is constructed by ensuring that it covers the same area as LS DR9, with a magnitude limit of $m_z \leq 21.0$ and a redshift interval of $[0.0, 1.0]$. Additionally, we apply bright-star masking to the MGRS. Relative to the original north galactic cap of MGRS, the galaxy count post masking decreases by 2.8 million, and the sky coverage shrinks by $\rm 476\ deg^2$. Figure \ref{LFJiutian} displays the LFs from our MGRS (MGRS-simu, depicted as dark brown open circles with error bars) alongside the observational data from Y1-z19.0 (magenta) and SV3-z19.0 (blue), showing a good match, which is anticipated due to the use of the abundance matching technique. 

In order to account for observational effects in redshift measurements, two additional redshifts are assigned to each galaxy in addition to the true MGRS-simu. The first of these is the MGRS-spec, which takes into account the redshift error of the DESI spectroscopic observation at about 35 km/s and the peculiar velocity of the galaxy. The other is the MGRS-photo, which includes a photoz error with a Gaussian distribution as described in \citet{yang2021extended}, $\sigma_{\rm z}$=(0.01+0.015z)*(1+z). To best mimic Y1-BGS, for galaxies with $m_z<19.0$, we randomly choose 45\% of them to keep the spectroscopic redshifts. 

The light brown and orange open circles in Figure \ref{LFJiutian} represent the LFs obtained from the MGRS-spec and MGRS-photo, respectively. All measurements based on MGRS agree, even for luminosities of $L \sim 10^{6} h^{-1}L_{\odot}$. 
The MGRS-simu and MGRS-spec are almost indistinguishable in all redshift bins, while the MGRS-photo displays a slightly smoother trend at the faint end in the lowest redshift bin, which is caused by the Gaussian photo-z error we applied. The behavior of MGRS-photo suggests that a pure Gaussian photo error cannot accurately reproduce the decreasing behavior of the observed LFs of the half-spectroscopic sample at the faint end, as seen in Figure \ref{dr9plots}. 

To delve deeper into the cause of the decline in LFs at the faint end due to the actual photo-z distributions observed in the DESI LS, we examine the redshift distribution of galaxies with an apparent magnitude of $18.5<r<19.5$, as the faint end LFs are mainly contributed by these galaxies. As depicted in Figure \ref{fig:zdist2} in the Appendix \ref{appendix:B}, in comparison to the spectroscopic redshift distribution, galaxies in the lowest redshift peaks at approximately $z=0$ exhibit a non-Gaussian photo-z distribution. Specifically, their photo-z distribution leans towards higher redshifts. Consequently, the observed decrease in the LFs/SMFs in the DESI data is primarily due to the skewed nature of the photo-z in the LS DR9.

\begin{figure*}[!ht]
\centering
\includegraphics[width=0.9\textwidth]{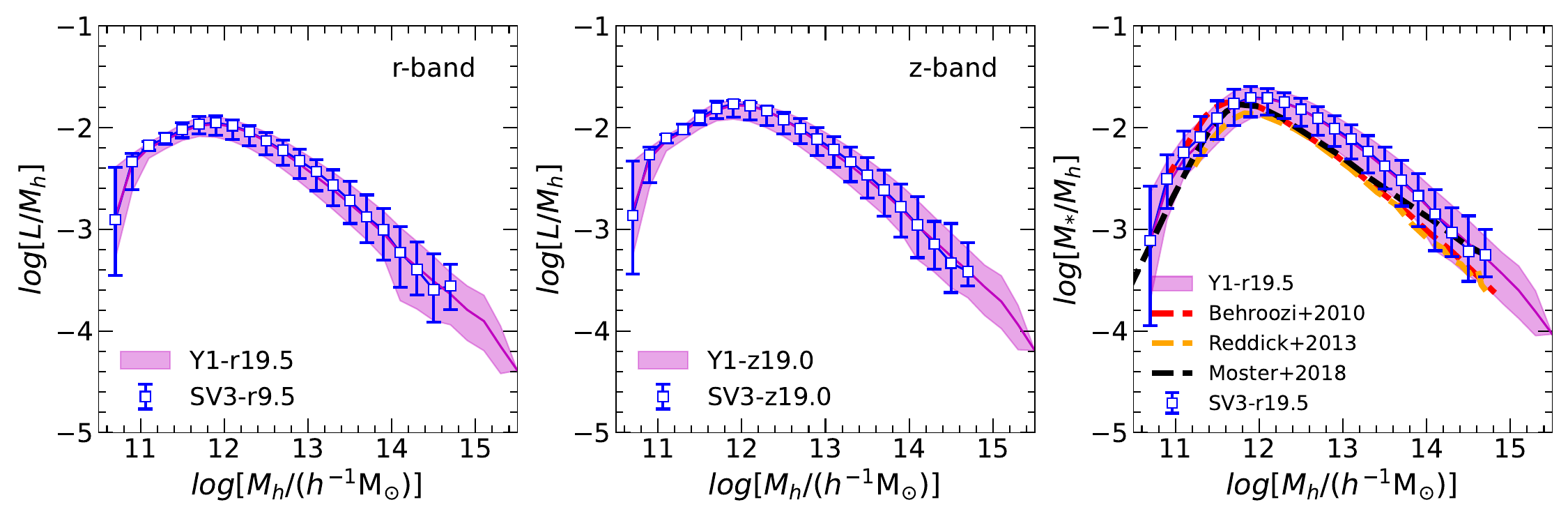}
\caption{Central galaxy luminosity (stellar mass) v.s. halo mass relations. The magenta-shaded regions are results obtained from the Y1-BGS sub-sample. Blue squares with error bars are obtained from the SV3-BGS sub-sample. The shaded regions and error bars represent 1-$\sigma$ (68$\%$) of scatters. Results from previous studies \citep{Behroozi2013, Reddick2013, Moster2018} are also shown in the far right panel for comparison. }\label{abmatching_galhalo}
\end{figure*}

\begin{figure*}[!ht]
\centering
\includegraphics[width=0.9\textwidth]{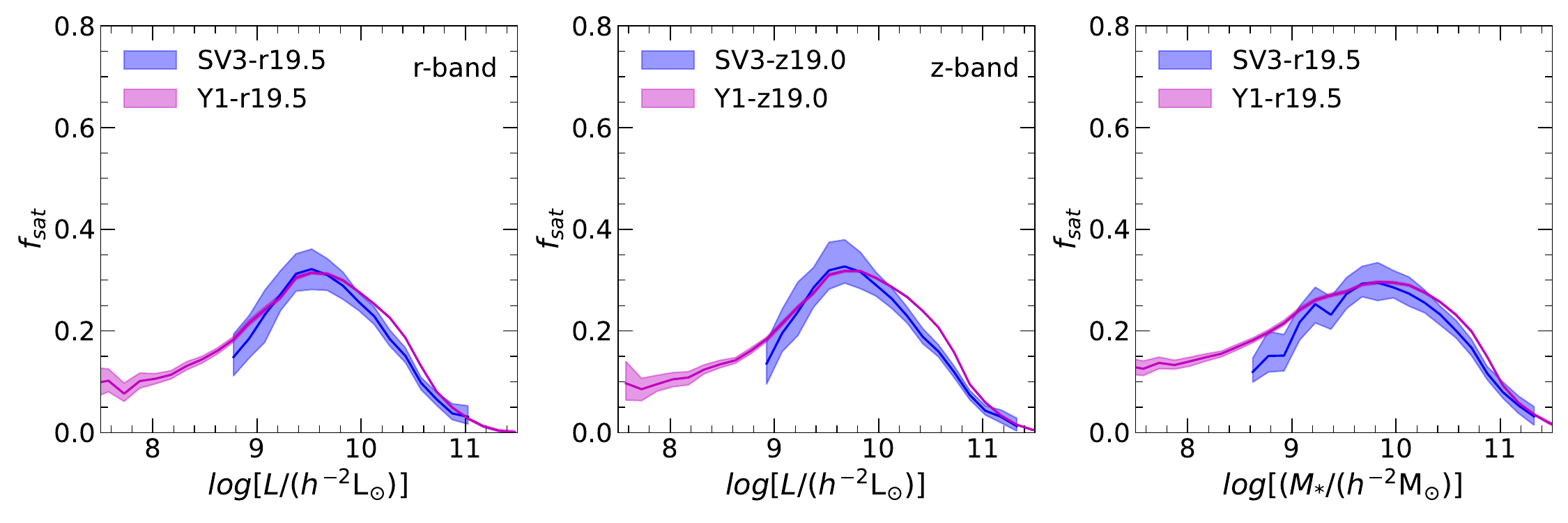}
\caption{Fraction of satellite galaxies for Y1-BGS and SV3-BGS sub-samples as a function of luminosity (left and middle panels for $r$-band and $z$-band, respectively) and stellar mass (right panel) using $V_{max}$ method. The error bars represent 3-$\sigma$ (99.8\%) of the scatters. Note that for each luminosity bin, we removed data points whose total galaxy numbers are less than 600. \label{fsat}}
\end{figure*}

\begin{figure*}[!ht]
\centering
\includegraphics[width=1\textwidth]{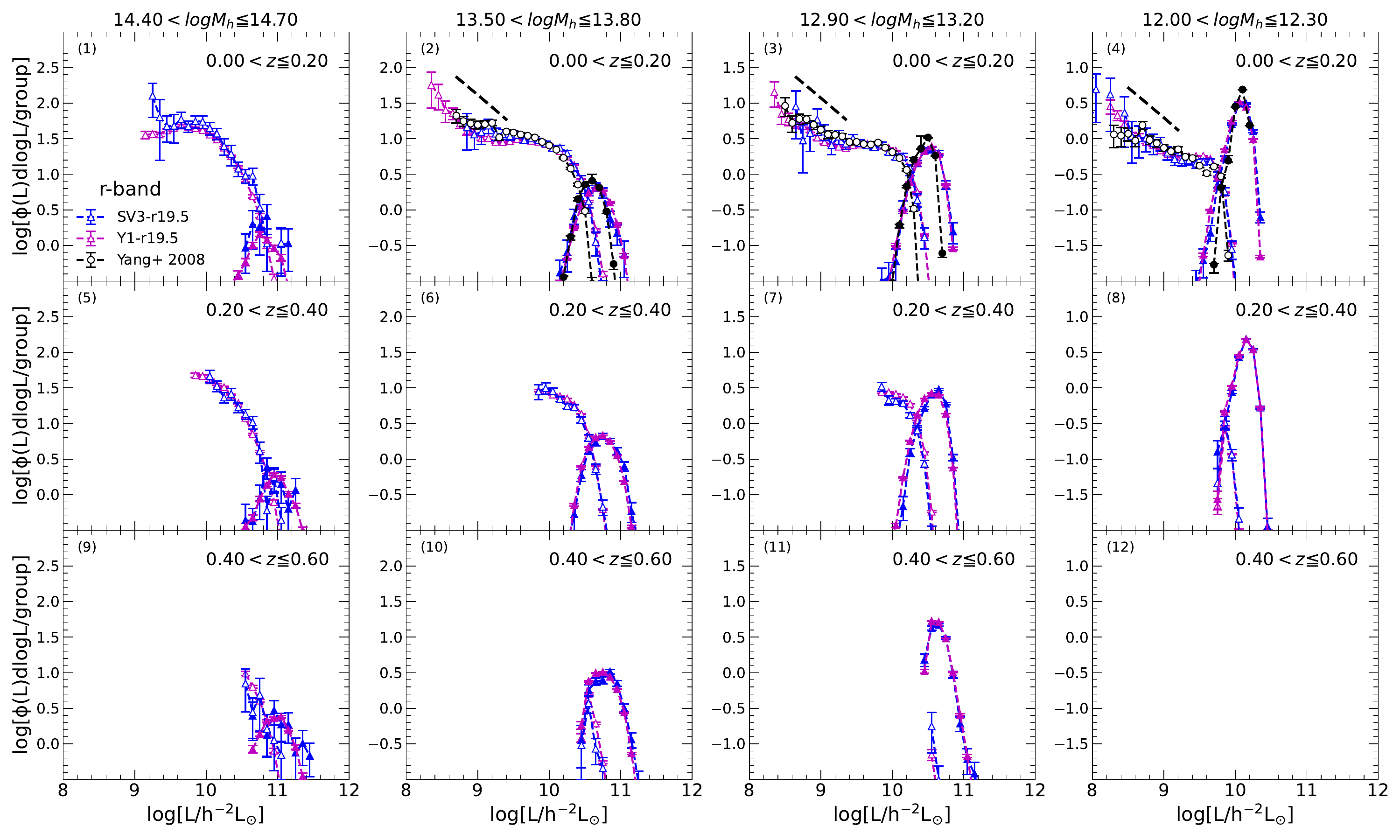}
\caption{CLFs ($r$-band) obtained from Y1-r19.5 (magenta triangles) and SV3-r19.5 (blue ones) sub-samples. Each column corresponds to a different halo mass bin and each row stands for a different redshift bin. The error bars are obtained from the standard deviation of 200 bootstrap resamplings. We removed data points whose galaxy numbers are less than 10 in each luminosity bin of the Y1-BGS sub-sample. The black dashed short strlines in the upper panels illustrate the slope of the subhalo mass function. Results from \citet{yang2008galaxy} are also shown in the lowest redshift bin for comparison. \label{CLFrband}}
\end{figure*}

\begin{figure*}[!ht]
\centering
\includegraphics[width=1\textwidth]{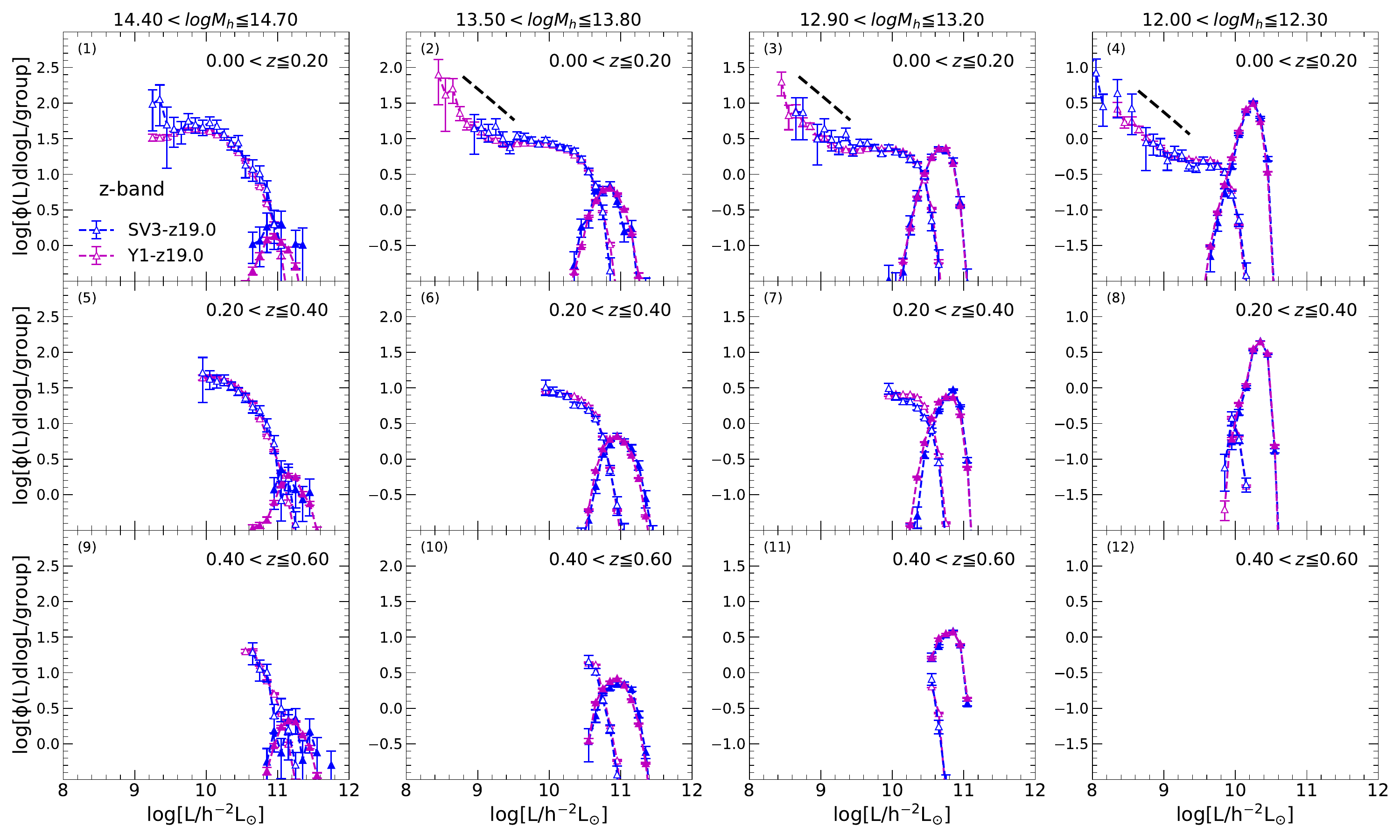}
\caption{CLFs ($z$-band) obtained from Y1-z19.0 (magenta triangles) and SV3-z19.0 (blue ones) sub-samples. Similarly with CLFs ($r$-band), each column corresponds to a different halo mass bin and each row stands for a different redshift bin. The error bars are obtained from the standard deviation of 200 bootstrap resamplings. We removed data points whose galaxy numbers are less than 10 in each luminosity bin of the Y1-BGS sub-sample. The black dashed short lines in the upper panels illustrate the slope of the subhalo mass function. \label{CLFzband}}
\end{figure*}

\begin{figure*}[!ht]
\centering
\includegraphics[width=1\textwidth]{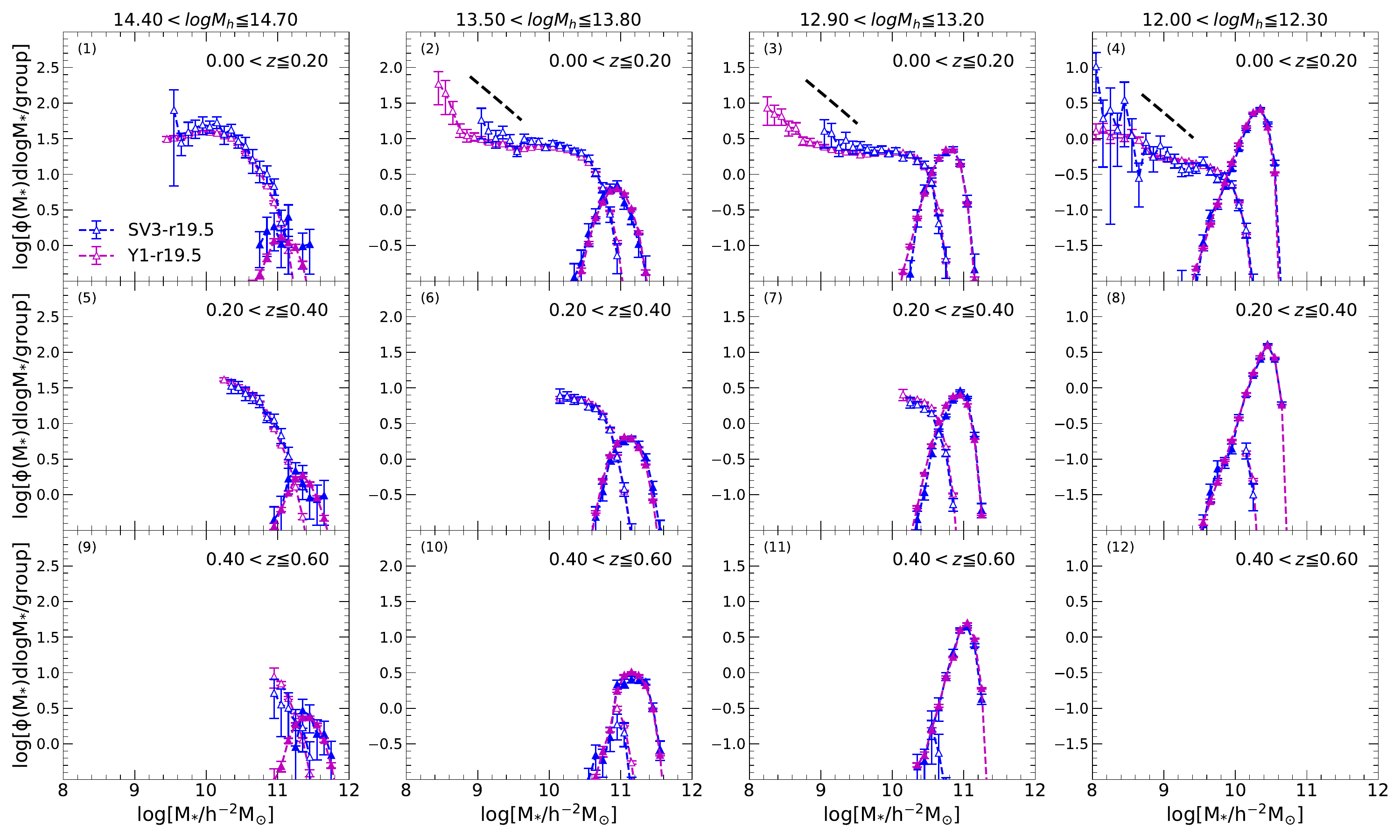}
\caption{CSMFs obtained from the combination of Y1-r19.5 (magenta triangles) and SV3-r19.5 (blue ones) sub-samples. Similarly with CLFs above, each column corresponds to a different halo mass bin and each row stands for a different redshift bin. The error bars are obtained from the standard deviation of 200 bootstrap resamplings. We removed data points whose galaxy numbers are less than 10 in each stellar mass bin of the Y1-BGS sub-sample. The black dashed short lines in the upper panels illustrate the slope of the subhalo mass function. \label{CSMF}}
\end{figure*}


\subsection{Comparing the true and measured CLFs}
\label{sec:result_Jiutian}

The CLF describes the average number of galaxies as a function of galaxy luminosity in the dark matter halo of a given mass, which plays an essential role in understanding how galaxies form and distribute in dark matter halos. 
In this sub-section, we assess the impact of the group finder on the CLF measurements by comparing the results from MGRS-spec and MGRS-photo 
to the true input values (MGRS-simu). To do this, we apply the same extended group finder from \citet{yang2021extended} to identify groups in our LS DR9 MGRS. We then construct two versions of mock group catalogs, one for the MGRS-spec and the other for MGRS-photo. We used MGRS-spec, MGRS-photo, and MGRS-simu to represent the results measured from the two mock group catalogs and the true input values, respectively. 

The CLF is measured differently from the LF, which is normalized by the number of groups in the redshift range
instead of the cosmic volume. 
The same critical step of determining the maximum redshift ${\rm z}_{\rm max}$ is still necessary. 
To calculate the CLF, galaxies and groups in the redshift range $[{\rm z}_1, {\rm z}_2]$ are first selected. For a galaxy with a specific absolute magnitude, the maximum redshift ${\rm z}_{\rm max}$ is then determined. The effective number of groups is calculated by counting the number of groups within the redshift range $[{\rm z}_1, {\rm min}({\rm z}_2,{\rm z}_{\rm max})]$. Figure \ref{CLFJiutian} shows the CLFs in the $z$ band obtained from the Jiutian MGRS in different halo mass and redshift bins. The redshift range is from $[0.0,0.2]$, $[0.2,0.4]$, and $[0.4,0.6]$ from the top to bottom panels, respectively. The halo mass bins are $[10^{14.4},10^{14.7}]$, $[10^{13.5},10^{13.8}]$, $[10^{12.9},10^{13.2}]$, and $[10^{12.0},10^{12.3}]$ (unit of $h^{-1}M_{\odot}$) from left to right, respectively.

We find that the central galaxy CLFs in both the MGRS-spec and MGRS-photo are accurately reproduced in halos with masses greater than $10^{13} \msunh$, regardless of redshift. The median values of these samples are similar to those of MGRS-simu in less massive halos, however, the scatter is slightly underestimated due to the estimation of the halo mass based on abundance matching \citep{yang2005halo}. \citet{Xu2023} recently reported that the group finder could create an artificial double-peak profile in the central CLF for a shallow survey. They found that the brighter component of the double-peak is largely contributed by the groups with a single-member galaxy (see Appendix \ref{appendix:A} for a more detailed discussion). Such artifacts can be greatly reduced by using a deeper survey. We do not observe this feature, which manifests the validation of using a deeper survey for the group-finding process. In summary, the CLFs for central galaxies in DESI DR9 can be well recovered, except for scatter in small halos.

The CLFs of satellites derived from the MGRS-spec, MGRS-photo and MGRS-simu samples, with luminosity greater than $10^{8}\Lsunhh$, also show very good agreement within different halo mass bins and redshift ranges. Most of the data points agree well with each other within their 1-$\sigma$ error bars. The only slight difference we can see is in the lowest halo mass bin, where the CLF from MGRS-spec tends to be slightly underestimated compared to MGRS-simu, with a difference of approximately 0.05dex. 
These comparisons demonstrate that the satellite CLFs of either a pure spec-z sample or a mixed sample (half photo-z and half spec-z) can also be well recovered, at least for satellite galaxies with $L\geq 10^{8}\Lsunhh$. The group finder will not induce significant bias in the CLF measurements in the DESI observations.


\section{Conditional luminosity and stellar mass functions}
\label{sec:CLF&CSMF}

With all of the above preparations, we set out to measure the CLFs and CSMFs from the DESI galaxy observational samples. 


\subsection{Global properties}
\label{sec:Ab_mat}

Before we move forward, we will present two sets of measurements of the global characteristics of groups that are essential components of the CLF and HOD theoretical framework:
(1) the luminosity (or stellar mass) of the central galaxy - halo mass relation, and (2) the satellite fraction, $f_{\rm sat}$.

The central galaxy luminosity (or stellar mass) - halo mass relation is a key factor in understanding how galaxies form and evolve in dark matter halos. We show in Figure \ref{abmatching_galhalo} the central galaxy - halo mass ratios
obtained from the LS DR9 group catalogs within $0 \le z\le 0.6$ for $r$ band luminosity, $z$ band luminosity and stellar mass in the left, middle, and right panels, respectively.  The magenta-shaded region represents the Y1-BGS subsample, and the blue squares with error bars correspond to the SV3-BGS subsample. There is no significant difference between the spec-z subsample (SV3-BGS) and the half spec-z subsample (Y1-BGS). This is in agreement with the results of Section~\ref{sec:Jiutian}, which suggests that the photo-z error has a negligible effect on the CLFs. Generally, the luminosity (stellar mass) of the centrals has the largest luminosity (stellar mass) to halo mass ratio in halos with mass $\sim 10^{12}\msunh$, then drops at the ends of low mass and high mass. This is consistent with previous findings, which are also shown in the right panel for comparison \citep{Behroozi2013,Reddick2013,Moster2018}, and is probably due to a combination of AGN feedback at the massive end and supernova feedback at the low mass end \citep[][]{yang2008galaxy}.

In the HOD framework, the satellite fraction $f_{\rm sat}$ is considered one of the most important quantities in modeling the galaxy correlation functions. $f_{\rm sat}$ is defined as the ratio between the number of satellites and the number of all galaxies at fixed galaxy luminosity or the stellar mass bin. Since each halo only contains one central galaxy, the large-scale clustering of galaxies at fixed luminosity is significantly impacted by the fraction of the satellite galaxies. 

The $f_{\rm sat}$ functions obtained from the SV3-BGS and Y1-BGS sub-samples are displayed in Figure \ref{fsat} for redshift range $0 \le z\le 0.6$. The fraction of satellites is calculated by dividing the number of satellites by the total number of galaxies in a given luminosity or stellar mass bin  normalized by the $V_{\rm max}$ factor. The error bars represent 3-$\sigma$ confidence levels obtained from 200 bootstrap resamplings. For accuracy, only results from luminosity bins with at least 600 galaxies are shown, which leads to a truncation of the SV3-BGS sub-sample at the faint end. The satellite fraction increases from close to 10\% at the faint end to a maximum of around 30\% at $L \sim 10^{9.5}\Lsunhh$ and then decreases to zero for the most luminous or massive galaxies. This 10\% level of the satellite fraction in the faint end (low mass) is in agreement with the satellite-to-all subhalo fraction at the low mass end in our Jiutian simulations. However, it should be noted that this fraction might vary on the basis of simulation resolution, the techniques used for subhalo identification, and whether disrupted subhalos are included. From an observational standpoint, our group finder only detects the brightest group galaxies (BGGs). \citet{Skibba2011} noted that a specific fraction of central galaxies are not BGGs. Consequently, the satellite fraction estimated by our group finder can be somewhat underestimated. The results in the three panels demonstrate that $f_{\rm sat}$ is independent of the choice of galaxy stellar mass and luminosity, as well as the bands of luminosity.


\subsection{CLFs measured from DESI observations}
\label{sec:CLF}

In this sub-section, we directly measure the CLFs from the DESI observational data in multiple halo mass bins. As tested in Section~\ref{sec:Jiutian}, the CLFs can be reliably measured from both the spectroscopic and combined redshift data. However, the available spectroscopic sample in this study, SV3-BGS, only covers about $\rm 133\ deg^2$. Therefore, it can only provide CLF measurements for relatively bright galaxies. The Y1-BGS sub-sample, on the other hand, covers a much larger area of the sky and has a spectroscopic redshift completeness of $\sim 50\%$. According to Section \ref{sec:result_Jiutian}, we have demonstrated that the mixed sample, with the same spectral completeness as Y1-BGS, can also provide reliable CLF for luminosities $L\geq 10^8\Lsunhh$. We present the direct CLF results of SV3-BGS and Y1-BGS sub-samples. Here, a small correction factor obtained from Section \ref{sec:LF} is applied to the Y1-BGS sub-sample. As we focus on relatively bright luminosity ranges ($L\ga 10^8\Lsunhh$), even without such a correction, general trends will remain almost unchanged.

Figure \ref{CLFrband} shows the CLFs measured from the $r$ band in different mass and redshift bins. Compared with those in Section \ref{sec:result_Jiutian}, we find that the CLFs of DESI show a similar behavior to those obtained from the MGRS, in that Y1-BGS and SV3-BGS show very similar results. On the other hand, due to the much larger sky coverage of Y1-BGS, its CLFs in general show much better statistics and smaller error bars.  

For comparison, we also show the $r$-band CLFs measured from SDSS observations in \citet{yang2008galaxy} with black circles in the top panels of Figure ~\ref{CLFrband}.
In this instance, we present results only for the three lower halo mass bins since their most massive bin covers a broader range of halo masses than ours. Generally, our findings are consistent with \citet{yang2008galaxy}. However, thanks to the DESI deeper imaging and spectroscopic surveys, our CLF measurements are able to reach much fainter end and show a clear upturn. Quite interestingly, such kind of enhancement in the faint end CLFs and CSMFs was already reported in previous works \citep[e.g.][]{Rodr2013, Lan2016, Meng2023} using SDSS groups. By combing  SDSS imaging data and SDSS group catalogs, \citet{Lan2016} found that the satellite CLFs of SDSS groups at redshift $0.01 \sim 0.05$ display a steep upturn at $L\la 10^{9}\Lsunhh$ for all halo masses, mainly contributed by red galaxies. 

Here, thanks to the much deeper DESI observations, we are able to obtain more reliable CLF measurements. Interestingly, we found that the slope of the upturn at $L\la 10^{9}\Lsunhh$ is rather steep ($\alpha \sim -1\pm0.3$). The faint end slope of the CLFs is in general agreement with that of the subhalo mass function, which is indicated by the black dashed line in the top panels, except for the most massive bin. This suggests that galaxies may have a roughly constant star formation efficiency in low-mass subhalos. 
According to Figure \ref{abmatching_galhalo}, a galaxy with a characteristic luminosity of $L\sim 10^{9}\Lsunhh$ lives in a halo with mass $M_h\sim 10^{11}\msunh$. Halos with mass lower than this critical mass tend to form stars with a roughly constant efficiency.

Figure \ref{CLFzband} shows the CLFs measured from $z$-band in different mass and redshift bins. Overall, the CLFs from $z$-band show similar trends as $r$-band, except that the ones in $z$-band are slightly shifted to brighter end.


\subsection{CSMFs measured from DESI observations}
\label{sec:CSMF}

The CSMF, $\Phi(M_{*} M_{h})$, is a key element in
modeling the evolution of galaxies. 
It describes the average number of galaxies as a function of galaxy stellar mass $M_{*}$ in a dark matter halo of a particular mass $M_{h}$. 
It is simpler and more common to use CSMF than CLF to 
access galaxy formation models because of 
the difficulty in converting mass to luminosity.
When evaluating the CSMF, the completeness of the galaxy sample in terms of stellar mass shall also be properly taken into account. For each galaxy, we count the number of groups within the maximum redshift $z_{\rm max}$ and redshift range to calculate the CSMFs. 
The CSMFs obtained from the DESI observations are shown in Figure \ref{CSMF}. The data points are taken from the SV3-BGS and Y1-BGS samples. The CSMFs have overall similarities with the CLF in terms of shape and features. 

The above direct CLF/CSMF measurements that covering much larger redshift and luminosity/stellar mass ranges, will be adopted in a subsequent study to understand the evolution of galaxies below redshift $z=0.6$.

\begin{table*}[!ht]
\centering
\caption{Values of the galaxy LFs and SMFs obtained from Y1-r19.5 sub-sample after being modified by correction factors ${\rm f}_{1}$ and ${\rm f}_{2}$, corresponding to Figure \ref{vs_chen}. The errorbars of the corrected values inherit the original ones.}
\resizebox{18cm}{!}{
\begin{tabular}{ccccccc}
\hline
\toprule
\multirow{2}{*}{\begin{tabular}[c]{@{}c@{}}$L$ or $M_{*}$ \\ 
$[\Lsunhh]$ ~~ $[\msunhh]$\end{tabular}} & \multicolumn{3}{c}{\begin{tabular}[c]{@{}c@{}}$\log\Phi(L)-rband$ \\  
$[h^{3}{\rm Mpc}^{-3}d\log L]$ \end{tabular}} & \multicolumn{3}{c}{\begin{tabular}[c]{@{}c@{}}$\log\Phi(M_{*})$ \\ 
$[h^{3}{\rm Mpc}^{-3}d\log M_{*}]$ \end{tabular}} \\ \cline{2-7} 
& Y1 & Y1+f1 & Y1+f1+f2  & Y1 & Y1+f1 & Y1+f1+f2      \\ \hline 
6.05 & $-0.8244_{-0.1281}^{+0.0988}$ & $0.06482_{-0.1281}^{+0.0988}$ & $-_{}^{}$  & $-1.0045_{-0.0895}^{+0.0742}$ & $-0.3407_{-0.0895}^{+0.0742}$ & $-_{}^{}$  \\
6.15 & $-0.9495_{-0.1312}^{+0.1006}$ & $-0.1161_{-0.1312}^{+0.1006}$ & $-_{}^{}$  & $-1.1299_{-0.0811}^{+0.0683}$ & $-0.5188_{-0.0811}^{+0.0683}$ & $-_{}^{}$  \\
6.25 & $-0.8858_{-0.0979}^{+0.0798}$ & $-0.1061_{-0.0979}^{+0.0798}$ & $-_{}^{}$  & $-0.8738_{-0.0822}^{+0.0691}$ & $-0.3129_{-0.0822}^{+0.0691}$ & $-_{}^{}$  \\
6.35 & $-0.7447_{-0.0616}^{+0.0539}$ & $-0.0168_{-0.0616}^{+0.0539}$ & $-_{}^{}$  & $-0.9529_{-0.0831}^{+0.0697}$ & $-0.4397_{-0.0831}^{+0.0697}$ & $-_{}^{}$  \\
6.45 & $-0.8351_{-0.0689}^{+0.0594}$ & $-0.1568_{-0.0689}^{+0.0594}$ & $-_{}^{}$  & $-1.0227_{-0.0773}^{+0.0656}$ & $-0.5547_{-0.0773}^{+0.0656}$ & $-_{}^{}$  \\
6.55 & $-0.8046_{-0.0559}^{+0.0495}$ & $-0.1739_{-0.0559}^{+0.0495}$ & $-_{}^{}$  & $-0.9623_{-0.0498}^{+0.0447}$ & $-0.5371_{-0.0498}^{+0.0447}$ & $-_{}^{}$  \\
6.65 & $-0.9882_{-0.0593}^{+0.0521}$ & $-0.4030_{-0.0593}^{+0.0521}$ & $-_{}^{}$  & $-1.1043_{-0.0446}^{+0.0404}$ & $-0.7195_{-0.0446}^{+0.0404}$ & $-_{}^{}$  \\
6.75 & $-1.0307_{-0.0404}^{+0.0369}$ & $-0.4890_{-0.0404}^{+0.0369}$ & $-_{}^{}$  & $-1.1221_{-0.0460}^{+0.0416}$ & $-0.7751_{-0.0460}^{+0.0416}$ & $-_{}^{}$  \\
6.85 & $-1.0179_{-0.0407}^{+0.0372}$ & $-0.5176_{-0.0407}^{+0.0372}$ & $-_{}^{}$  & $-1.2026_{-0.0307}^{+0.0287}$ & $-0.8911_{-0.0307}^{+0.0287}$ & $-_{}^{}$  \\
6.95 & $-1.0794_{-0.0364}^{+0.0335}$ & $-0.6185_{-0.0364}^{+0.0335}$ & $-_{}^{}$  & $-1.2648_{-0.0307}^{+0.0286}$ & $-0.9863_{-0.0307}^{+0.0286}$ & $-_{}^{}$  \\
7.05 & $-1.1274_{-0.0307}^{+0.0287}$ & $-0.7038_{-0.0307}^{+0.0287}$ & $-_{}^{}$  & $-1.3167_{-0.0276}^{+0.0259}$ & $-1.0687_{-0.0276}^{+0.0259}$ & $-_{}^{}$  \\
7.15 & $-1.2113_{-0.0330}^{+0.0307}$ & $-0.8229_{-0.0330}^{+0.0307}$ & $-_{}^{}$  & $-1.3448_{-0.0224}^{+0.0213}$ & $-1.1248_{-0.0224}^{+0.0213}$ & $-_{}^{}$  \\
7.25 & $-1.2318_{-0.0262}^{+0.0247}$ & $-0.8766_{-0.0262}^{+0.0247}$ & $-_{}^{}$  & $-1.3473_{-0.0186}^{+0.0178}$ & $-1.1528_{-0.0186}^{+0.0178}$ & $-_{}^{}$  \\
7.35 & $-1.2753_{-0.0226}^{+0.0215}$ & $-0.9512_{-0.0226}^{+0.0215}$ & $-0.2555_{-0.0226}^{+0.0215}$  & $-1.3313_{-0.0169}^{+0.0163}$ & $-1.1600_{-0.0169}^{+0.0163}$ & $-0.4644_{-0.0169}^{+0.0163}$  \\
7.45 & $-1.3208_{-0.0187}^{+0.0180}$ & $-1.0258_{-0.0187}^{+0.0180}$ & $-0.3301_{-0.0187}^{+0.0180}$  & $-1.3178_{-0.0125}^{+0.0122}$ & $-1.1671_{-0.0125}^{+0.0122}$ & $-0.4715_{-0.0125}^{+0.0122}$  \\
7.55 & $-1.2930_{-0.0165}^{+0.0159}$ & $-1.0250_{-0.0165}^{+0.0159}$ & $-0.4627_{-0.0165}^{+0.0159}$  & $-1.3325_{-0.0110}^{+0.0108}$ & $-1.2000_{-0.0110}^{+0.0108}$ & $-0.6377_{-0.0110}^{+0.0108}$  \\
7.65 & $-1.2980_{-0.0119}^{+0.0116}$ & $-1.055_{-0.0119}^{+0.0116}$ & $-0.6023_{-0.0119}^{+0.0116}$  & $-1.3320_{-0.0090}^{+0.0088}$ & $-1.2152_{-0.0090}^{+0.0088}$ & $-0.7626_{-0.0090}^{+0.0088}$  \\
7.75 & $-1.3157_{-0.0125}^{+0.0122}$ & $-1.0955_{-0.0125}^{+0.0122}$ & $-0.7102_{-0.0125}^{+0.0122}$  & $-1.3237_{-0.0079}^{+0.0077}$ & $-1.2202_{-0.0079}^{+0.0077}$ & $-0.8349_{-0.0079}^{+0.0077}$  \\
7.85 & $-1.2772_{-0.0092}^{+0.0090}$ & $-1.0779_{-0.0092}^{+0.0090}$ & $-0.7678_{-0.0092}^{+0.0090}$  & $-1.3311_{-0.0065}^{+0.0064}$ & $-1.2385_{-0.0065}^{+0.0064}$ & $-0.9284_{-0.0065}^{+0.0064}$  \\
7.95 & $-1.2571_{-0.0073}^{+0.0072}$ & $-1.0765_{-0.0073}^{+0.0072}$ & $-0.8485_{-0.0073}^{+0.0072}$  & $-1.3357_{-0.0060}^{+0.0060}$ & $-1.2515_{-0.0060}^{+0.0060}$ & $-1.0234_{-0.0060}^{+0.0060}$  \\
8.05 & $-1.2581_{-0.0065}^{+0.0064}$ & $-1.0941_{-0.0065}^{+0.0064}$ & $-0.9394_{-0.0065}^{+0.0064}$  & $-1.3403_{-0.0050}^{+0.0049}$ & $-1.2619_{-0.0050}^{+0.0049}$ & $-1.1071_{-0.0050}^{+0.0049}$  \\
8.15 & $-1.2655_{-0.0051}^{+0.0050}$ & $-1.1162_{-0.0051}^{+0.0050}$ & $-1.0238_{-0.0051}^{+0.0050}$  & $-1.3625_{-0.0055}^{+0.0055}$ & $-1.2876_{-0.0055}^{+0.0055}$ & $-1.1952_{-0.0055}^{+0.0055}$  \\
8.25 & $-1.2686_{-0.0048}^{+0.0048}$ & $-1.132_{-0.0048}^{+0.0048}$ & $-1.0961_{-0.0048}^{+0.0048}$  & $-1.3818_{-0.0039}^{+0.0039}$ & $-1.3079_{-0.0039}^{+0.0039}$ & $-1.2720_{-0.0039}^{+0.0039}$  \\
8.35 & $-1.2770_{-0.0041}^{+0.0041}$ & $-1.1509_{-0.0041}^{+0.0041}$ & $-1.1521_{-0.0041}^{+0.0041}$  & $-1.4013_{-0.0040}^{+0.0040}$ & $-1.3259_{-0.0040}^{+0.0040}$ & $-1.3272_{-0.0040}^{+0.0040}$  \\
8.45 & $-1.2924_{-0.0037}^{+0.0036}$ & $-1.2924_{-0.0037}^{+0.0036}$ & $-1.2937_{-0.0037}^{+0.0036}$  & $-1.4228_{-0.0037}^{+0.0036}$ & $-1.4228_{-0.0037}^{+0.0036}$ & $-1.4241_{-0.0037}^{+0.0036}$  \\
8.55 & $-1.3251_{-0.0030}^{+0.0030}$ & $-1.3251_{-0.0030}^{+0.0030}$ & $-1.3251_{-0.0030}^{+0.0030}$  & $-1.4642_{-0.0794}^{+0.0671}$ & $-1.4642_{-0.0794}^{+0.0671}$ & $-1.4642_{-0.0794}^{+0.0671}$  \\
8.65 & $-1.3700_{-0.0022}^{+0.0022}$ & $-1.3700_{-0.0022}^{+0.0022}$ & $-1.3700_{-0.0022}^{+0.0022}$  & $-1.4986_{-0.0029}^{+0.0028}$ & $-1.4986_{-0.0029}^{+0.0028}$ & $-1.4986_{-0.0029}^{+0.0028}$  \\
8.75 & $-1.4067_{-0.0022}^{+0.0022}$ & $-1.4067_{-0.0022}^{+0.0022}$ & $-1.4067_{-0.0022}^{+0.0022}$  & $-1.5257_{-0.0027}^{+0.0026}$ & $-1.5257_{-0.0027}^{+0.0026}$ & $-1.5257_{-0.0027}^{+0.0026}$  \\
8.85 & $-1.4502_{-0.0019}^{+0.0019}$ & $-1.4502_{-0.0019}^{+0.0019}$ & $-1.4502_{-0.0019}^{+0.0019}$  & $-1.5697_{-0.0020}^{+0.0020}$ & $-1.5697_{-0.0020}^{+0.0020}$ & $-1.5697_{-0.0020}^{+0.0020}$  \\
8.95 & $-1.4815_{-0.0019}^{+0.0019}$ & $-1.4815_{-0.0019}^{+0.0019}$ & $-1.4815_{-0.0019}^{+0.0019}$  & $-1.6000_{-0.0019}^{+0.0019}$ & $-1.6000_{-0.0019}^{+0.0019}$ & $-1.6000_{-0.0019}^{+0.0019}$  \\
9.05 & $-1.5181_{-0.0017}^{+0.0017}$ & $-1.5181_{-0.0017}^{+0.0017}$ & $-1.5181_{-0.0017}^{+0.0017}$  & $-1.6228_{-0.0020}^{+0.0020}$ & $-1.6228_{-0.0020}^{+0.0020}$ & $-1.6228_{-0.0020}^{+0.0020}$  \\
9.15 & $-1.5494_{-0.0015}^{+0.0015}$ & $-1.5494_{-0.0015}^{+0.0015}$ & $-1.5494_{-0.0015}^{+0.0015}$  & $-1.6316_{-0.0016}^{+0.0016}$ & $-1.6316_{-0.0016}^{+0.0016}$ & $-1.6316_{-0.0016}^{+0.0016}$  \\
9.25 & $-1.5641_{-0.0012}^{+0.0012}$ & $-1.5641_{-0.0012}^{+0.0012}$ & $-1.5641_{-0.0012}^{+0.0012}$  & $-1.6516_{-0.0016}^{+0.0016}$ & $-1.6516_{-0.0016}^{+0.0016}$ & $-1.6516_{-0.0016}^{+0.0016}$  \\
9.35 & $-1.5750_{-0.0013}^{+0.0012}$ & $-1.5750_{-0.0013}^{+0.0012}$ & $-1.5750_{-0.0013}^{+0.0012}$  & $-1.6771_{-0.0023}^{+0.0023}$ & $-1.6771_{-0.0023}^{+0.0023}$ & $-1.6771_{-0.0023}^{+0.0023}$  \\
9.45 & $-1.5771_{-0.0009}^{+0.0009}$ & $-1.5771_{-0.0009}^{+0.0009}$ & $-1.5771_{-0.0009}^{+0.0009}$  & $-1.7021_{-0.0012}^{+0.0012}$ & $-1.7021_{-0.0012}^{+0.0012}$ & $-1.7021_{-0.0012}^{+0.0012}$  \\
9.55 & $-1.5974_{-0.0008}^{+0.0008}$ & $-1.5974_{-0.0008}^{+0.0008}$ & $-1.5974_{-0.0008}^{+0.0008}$  & $-1.7217_{-0.0011}^{+0.0011}$ & $-1.7217_{-0.0011}^{+0.0011}$ & $-1.7217_{-0.0011}^{+0.0011}$  \\
9.65 & $-1.6253_{-0.0008}^{+0.0008}$ & $-1.6253_{-0.0008}^{+0.0008}$ & $-1.6253_{-0.0008}^{+0.0008}$  & $-1.7293_{-0.0009}^{+0.0009}$ & $-1.7293_{-0.0009}^{+0.0009}$ & $-1.7293_{-0.0009}^{+0.0009}$  \\
9.75 & $-1.6601_{-0.0006}^{+0.0006}$ & $-1.6601_{-0.0006}^{+0.0006}$ & $-1.6601_{-0.0006}^{+0.0006}$  & $-1.7453_{-0.0008}^{+0.0008}$ & $-1.7453_{-0.0008}^{+0.0008}$ & $-1.7453_{-0.0008}^{+0.0008}$  \\
9.85 & $-1.7123_{-0.0007}^{+0.0007}$ & $-1.7123_{-0.0007}^{+0.0007}$ & $-1.7123_{-0.0007}^{+0.0007}$  & $-1.7616_{-0.0008}^{+0.0008}$ & $-1.7616_{-0.0008}^{+0.0008}$ & $-1.7616_{-0.0008}^{+0.0008}$  \\
9.95 & $-1.7757_{-0.0007}^{+0.0007}$ & $-1.7757_{-0.0007}^{+0.0007}$ & $-1.7757_{-0.0007}^{+0.0007}$  & $-1.7861_{-0.0008}^{+0.0008}$ & $-1.7861_{-0.0008}^{+0.0008}$ & $-1.7861_{-0.0008}^{+0.0008}$  \\
10.05 & $-1.8442_{-0.0007}^{+0.0007}$ & $-1.8442_{-0.0007}^{+0.0007}$ & $-1.8442_{-0.0007}^{+0.0007}$  & $-1.8118_{-0.0007}^{+0.0007}$ & $-1.8118_{-0.0007}^{+0.0007}$ & $-1.8118_{-0.0007}^{+0.0007}$  \\
10.15 & $-1.9324_{-0.0008}^{+0.0008}$ & $-1.9324_{-0.0008}^{+0.0008}$ & $-1.9324_{-0.0008}^{+0.0008}$  & $-1.8482_{-0.0008}^{+0.0008}$ & $-1.8482_{-0.0008}^{+0.0008}$ & $-1.8482_{-0.0008}^{+0.0008}$  \\
10.25 & $-2.0419_{-0.0009}^{+0.0009}$ & $-2.0419_{-0.0009}^{+0.0009}$ & $-2.0419_{-0.0009}^{+0.0009}$  & $-1.8975_{-0.0008}^{+0.0008}$ & $-1.8975_{-0.0008}^{+0.0008}$ & $-1.8975_{-0.0008}^{+0.0008}$  \\
10.35 & $-2.1730_{-0.0010}^{+0.0010}$ & $-2.1730_{-0.0010}^{+0.0010}$ & $-2.1730_{-0.0010}^{+0.0010}$  & $-1.9553_{-0.0010}^{+0.0010}$ & $-1.9554_{-0.0010}^{+0.0010}$ & $-1.9554_{-0.0010}^{+0.0010}$  \\
10.45 & $-2.3346_{-0.0012}^{+0.0012}$ & $-2.3346_{-0.0012}^{+0.0012}$ & $-2.3346_{-0.0012}^{+0.0012}$  & $-2.0434_{-0.0010}^{+0.0010}$ & $-2.0434_{-0.0010}^{+0.0010}$ & $-2.0434_{-0.0010}^{+0.0010}$  \\
10.55 & $-2.5286_{-0.0015}^{+0.0015}$ & $-2.5286_{-0.0015}^{+0.0015}$ & $-2.5286_{-0.0015}^{+0.0015}$  & $-2.1502_{-0.0011}^{+0.0011}$ & $-2.1502_{-0.0011}^{+0.0011}$ & $-2.1502_{-0.0011}^{+0.0011}$  \\
10.65 & $-2.7615_{-0.0025}^{+0.0025}$ & $-2.7615_{-0.0025}^{+0.0025}$ & $-2.7615_{-0.0025}^{+0.0025}$  & $-2.2771_{-0.0012}^{+0.0012}$ & $-2.2771_{-0.0012}^{+0.0012}$ & $-2.2771_{-0.0012}^{+0.0012}$  \\
10.75 & $-3.0362_{-0.0031}^{+0.0031}$ & $-3.0362_{-0.0031}^{+0.0031}$ & $-3.0362_{-0.0031}^{+0.0031}$  & $-2.4380_{-0.0013}^{+0.0013}$ & $-2.4380_{-0.0013}^{+0.0013}$ & $-2.4380_{-0.0013}^{+0.0013}$  \\
10.85 & $-3.3658_{-0.0048}^{+0.0048}$ & $-3.3658_{-0.0048}^{+0.0048}$ & $-3.3658_{-0.0048}^{+0.0048}$  & $-2.6302_{-0.0019}^{+0.0019}$ & $-2.6302_{-0.0019}^{+0.0019}$ & $-2.6302_{-0.0019}^{+0.0019}$  \\
10.95 & $-3.7564_{-0.0074}^{+0.0073}$ & $-3.7564_{-0.0074}^{+0.0073}$ & $-3.7564_{-0.0074}^{+0.0073}$  & $-2.8699_{-0.0023}^{+0.0023}$ & $-2.8699_{-0.0023}^{+0.0023}$ & $-2.8699_{-0.0023}^{+0.0023}$  \\
11.05 & $-4.1972_{-0.0131}^{+0.0127}$ & $-4.1972_{-0.0131}^{+0.0127}$ & $-4.1972_{-0.0131}^{+0.0127}$  & $-3.1467_{-0.0035}^{+0.0034}$ & $-3.1467_{-0.0035}^{+0.0034}$ & $-3.1467_{-0.0035}^{+0.0034}$  \\
11.15 & $-4.7686_{-0.0236}^{+0.0224}$ & $-4.7686_{-0.0236}^{+0.0224}$ & $-4.7686_{-0.0236}^{+0.0224}$  & $-3.4861_{-0.0051}^{+0.0050}$ & $-3.4861_{-0.0051}^{+0.0050}$ & $-3.4861_{-0.0051}^{+0.0050}$  \\
11.25 & $-5.4712_{-0.0522}^{+0.0466}$ & $-5.4712_{-0.0522}^{+0.0466}$ & $-5.4712_{-0.0522}^{+0.0466}$  & $-3.8765_{-0.0079}^{+0.0077}$ & $-3.8765_{-0.0079}^{+0.0077}$ & $-3.8765_{-0.0079}^{+0.0077}$  \\
11.35 & $-6.2598_{-0.1451}^{+0.1085}$ & $-6.2598_{-0.1451}^{+0.1085}$ & $-6.2598_{-0.1451}^{+0.1085}$  & $-4.3476_{-0.0138}^{+0.0134}$ & $-4.3476_{-0.0138}^{+0.0134}$ & $-4.3476_{-0.0138}^{+0.0134}$  \\
11.45 & $-6.7732_{-0.3395}^{+0.1882}$ & $-6.7732_{-0.3395}^{+0.1882}$ & $-6.7732_{-0.3395}^{+0.1882}$  & $-4.9537_{-0.0281}^{+0.0264}$ & $-4.9537_{-0.0281}^{+0.0264}$ & $-4.9537_{-0.0281}^{+0.0264}$  \\
11.55 & $-_{ }^{ }$ & $-_{ }^{ }$ & $-_{ }^{ }$  & $-5.6075_{-0.0586}^{+0.0516}$ & $-5.6075_{-0.0586}^{+0.0516}$ & $-5.6075_{-0.0586}^{+0.0516}$  \\
11.65 & $-_{ }^{ }$ & $-_{ }^{ }$ & $-_{ }^{ }$  & $-6.5013_{-0.1993}^{+0.1361}$ & $-6.5013_{-0.1993}^{+0.1361}$ & $-6.5013_{-0.1993}^{+0.1361}$  \\
11.75 & $-_{ }^{ }$ & $-_{ }^{ }$ & $-_{ }^{ }$  & $-7.0417_{-0.5120}^{+0.2285}$ & $-7.0417_{-0.5120}^{+0.2285}$ & $-7.0417_{-0.5120}^{+0.2285}$  \\
\hline
\end{tabular}}
\label{LF_SMF_result_aftcorr}
\end{table*}

\section{Discussion} 
\label{sec:Discussion}

\begin{figure}[!ht]
\centering
\includegraphics[width=0.42
\textwidth]{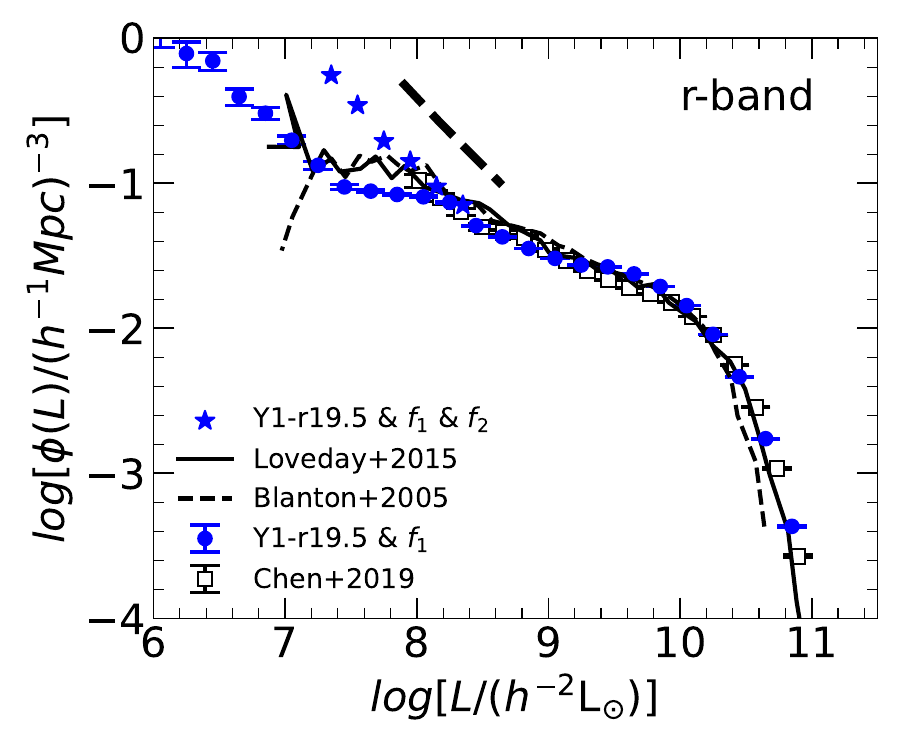}
\includegraphics[width=0.42
\textwidth]{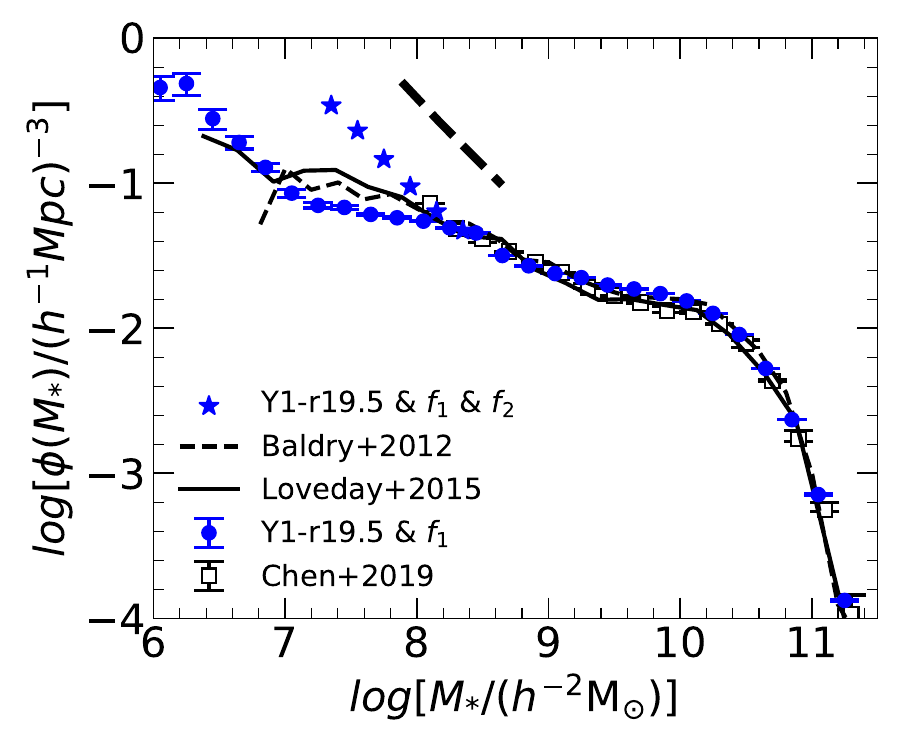}
\caption{LFs and SMFs of Y1-r19.5 sub-sample and those obtained by \citet{Chen2019} based on SDSS DR7. The blue symbols correspond to redshift ranges $[0.00,0.20]$, and the black open square with errorbas from \citet{Chen2019} is $[0.01, 0.12]$. ${\rm f}_{1}$ stand for the photoz correction factor and ${\rm f}_{2}$ is the cosmic variance correction factor. Values of data points are listed in Table \ref{LF_SMF_result_aftcorr}. The black short dashed line illustrates the slope of the low mass end of the halo mass function. Results from previous studies \citep[]{Loveday2015,Blanton2005,Baldry2012} are also shown for comparison.  \label{vs_chen} }
\end{figure}

In our measurements of the CLFs/CSMFs, we find a clear upturn at $L\la 10^{9}\Lsunhh$. However, this trend is not clearly pronounced in the LFs/SMFs of SV3-BGS shown in Figure \ref{dr9plots}, which exhibit a slight drop below $L\la 10^{8}\Lsunhh$. We set out to explore the clues for the potential discrepancy and find that the local void (LV) in our Universe is a possible reason \citep{Chen2019}. 
 
The Milky Way resides within a local void in the Universe. To mitigate the influence of this local void on the luminosity function  and stellar mass function  measurements, \citet{Chen2019} introduced a comprehensive correction approach using the ELUCID simulation \citep[e.g.][]{wang2014measuring, wang2016elucid, yang2018}, by comparing the LFs/SMFs from the SDSS region to those across the entire simulation box. Without this correction, the faint end of the galaxy luminosity functions derived from SDSS would be notably diminished.

To assess the constraining power of Y1-BGS in LFs at $L \ga 10^{8.2} h^{-1}L_{\odot}$, we compare our measurements of Y1-BGS that incorporate the photoz correction factor ${\rm f}_{1}$ from Section \ref{sec:LF&SMF} with those obtained from SDSS by \citet{Chen2019} in Figure \ref{vs_chen}. The galaxies in the two observations are both K-corrected to $z=0.1$ and the primary difference between these two samples is their redshift range, as $0<z<0.2$ in Y1-BGS, and $0<z<0.12$ in \citet{Chen2019}. We find that the LF results measured directly by Y1-BGS with the photoz correction factor ${\rm f}_{1}$ are in good agreement with those obtained from SDSS after LV correction. Y1-BGS, which has observed wide sky regions and is much deeper, should suffer from the LV effect in the much fainter luminosity ranges. The good agreement demonstrates that the LV correction factor obtained by \citet{Chen2019} in the range of $10^{8.2} h^{-2}L_{\odot}$ to $10^{9} h^{-2}L_{\odot}$ works remarkably well. Additionally, we compare our results using the correction factor ${\rm f}_{1}$ with earlier studies \citep[]{Loveday2015,Blanton2005,Baldry2012} as depicted in Figure \ref{vs_chen}, finding that the trends in the LFs and SMFs generally align.

Considering that the apparent magnitude limit of Y1-r19.5 is nearly two magnitudes fainter than \citet{Chen2019} which is 17.6, the impact of LV on the LFs might also appear in the two magnitudes fainter ranges, that is, below $L \la 10^{8.2} h^{-1}L_{\odot}$. 
In this section, we apply the same correction factor obtained by \citet{Chen2019} to our LFs and SMFs measurements at $L \la 10^{8.2} h^{-1}L_{\odot}$, shifted to the fainter end by 0.8dex in terms of luminosity or stellar mass. 
Our LF and SMF results with photo-z correction factor ${\rm f}_{1}$ are shown in Figure \ref{vs_chen} illustrated by blue dots with error bars. The results with the additional cosmic variance correction factor ${\rm f}_{2}$ are shown by blue star icons. 
It is evident that there is a considerable difference in the faint-end slope, with $\Delta \alpha \sim 0.5$ comparing the slope in the range of $10^{7.4} h^{-2}L_{\odot} \sim 10^{8.5} h^{-2}L_{\odot}$ with that of $L \ga 10^{8.5} h^{-2}L_{\odot}$. 

As a reference, we show in each panel of Figure \ref{vs_chen} the low mass end slope of the halo mass function using a black short dashed line.
Upon closer inspection, it is notable that the slopes of the LF and SMF match the typical halo mass function at the very faint end. This behavior is now in good agreement with our CLF measurements, in that for both total and satellite galaxies with luminosity/stellar mass less than $\sim 10^{9}\Lsunhh$ (or $\msunhh$), they show roughly consistent galaxy-halo connections.  In addition, it suggests that galaxy formation in low-mass halos could be still quite efficient, regardless of halo mass. This would be contrary to the standard theories of galaxy formation, which usually invoke stellar winds, supernova feedback, etc., to reduce star formation efficiency towards the low mass end.

It should be noted that our analysis did not incorporate surface brightness corrections for lower mass objects. \citet{Blanton2005} comprehensively discussed the issue of missing low surface brightness galaxies due to the SDSS pipeline. Nevertheless, with the deeper imaging capabilities and enhancements in the DESI LS pipeline, we anticipate that this problem will be less significant compared to SDSS. According to Figure 2 in \citet{Zou2019}, which illustrates the detection completeness of sources in the DESI LS compared to the COSMOS space observation, our BGS galaxies achieve a completeness of more than 97\%. 


\section{Summary} 
\label{sec:Conclusion}

In this paper, we leverage the two recent DESI observational sub-samples, SV3-BGS and Y1-BGS, to update the seed catalog of LS DR9 for our group finder. The SV3-BGS, despite the limited coverage of $\rm 133\ deg^2$, contains the most complete spectroscopic redshift data. In contrast, the Y1-BGS, with only half of the spectroscopic redshift completeness, achieves a sky coverage that is 90 times larger than SV3-BGS. We obtain a galaxy group catalog by applying the extended version of halo-based group finder \citep{yang2021extended} to this updated seed catalog. Based on the assessments using MGRS spec-z and photo-z mock galaxy samples constructed based on Jiutian simulation, we investigated the galaxy luminosity functions (LFs), stellar mass functions (SMFs), conditional luminosity functions (CLFs), and conditional stellar mass functions (CSMFs) in three redshift bins up to $z=0.6$. Our main results can be summarized as follows. 

\begin{enumerate}

\item We measure the galaxy LFs and SMFs in three different redshift bins. We find that utilization of photometric redshift in the mixed sub-sample will somewhat suppress the LFs and SMFs at the very faint/low mass end in the lowest redshift bin. 

\item We constructed an MGRS galaxy catalog based on Jiutian simulation using the LFs of Y1-BGS by applying the photo-z correction factor. To mimic the redshift completeness of SV3-BGS and Y1-BGS sub-samples, we have constructed two sets of redshifts, MGRS-spec and MGRS-photo. By applying the same group finder to the two sets of MGRSs, we use the resulting group catalogs to evaluate the reliability of the CLF measurements. Compared to the true values, the two MGRS samples demonstrate that the central galaxy CLFs can be accurately recovered using both spec-z and photo-z in all redshift and halo mass bins, except for the lowest halo mass bin. The CLFs of the satellite galaxies are slightly underestimated by approximately 0.05 dex using spec-z.

\item We derived the central luminosity (or stellar mass) - host halo mass relations and the satellite fraction based on the galaxy group catalogs constructed from SV3-BGS and Y1-BGS observational sub-samples, which extend down to a luminosity or stellar mass $\sim 10^{7.5}\Lsunhh$  ($\msunhh$). We found that the satellite fraction peaks at $\sim 10^{9.5}\Lsunhh$ at about 30\% level and decreases to 10\% at the low luminosity (or stellar mass) end.

\item Based on the validation of the group finder on our CLF measurements,  we provide our observational measurements of CLFs and CSMFs at $L > 10^{8}\Lsunhh$ ($\msunhh$) from SV3-BGS and Y1-BGS sub-samples in a wide halo mass range and three redshift bins. Our analysis reveals an upturn in the CLFs and CSMFs at the faint (or low stellar mass) end below $10^{9}\Lsunhh$ (or $\msunhh$). Remarkably, the slope of this upturn is in nice agreement with that of the subhalo mass functions. 

\item After taking into account the photo-z correction factor ${\rm f}_{1}$ and local void correction factor ${\rm f}_{2}$, the LFs and SMFs we obtained from DESI observation may also reveal a continuous upturn below $10^{9}\Lsunhh$ (or $\msunhh$), similar to those in the CLFs and CSMFs. The slope is in nice agreement with that of the halo mass function at the low mass end. 
\end{enumerate}

This study provides a comprehensive analysis of the LFs, SMFs, CLFs, and CSMFs of galaxies across a broad range of redshifts and halo mass bins, combining both observed and mock galaxy samples. These measurements also span large luminosity (or stellar mass) ranges of $\ga 10^{6.5} \Lsunhh$ (or $\msunhh$) at low redshift ${\rm z}\sim 0.1$ and $\ga 10^{10.5}\Lsunhh$ (or $\msunhh$) at higher redshift ${\rm z}\sim 0.5$.
The intriguing upturn feature in the faint (low mass) end of LFs/SMFs/CLFs/CSMFs carries significant implications for refining the CLF model. Moreover, it provides valuable insights into the formation and evolution mechanisms of galaxies in the very low mass halo. We will perform related investigations in a subsequent work.


\section*{Acknowledgements}
The authors thank Antonella Palmese and the anonymous referee for valuable comments that improved the presentation of this paper. 
This work is supported by the National Key R\&D Program of China (2023YFA1607800, 2023YFA1607804), the National Science Foundation of China (Nos. 11833005, 11890692, 11621303, 12141302), “the Fundamental Research Funds for the Central Universities”, 111 project No. B20019, and Shanghai Natural Science Foundation, grant No.19ZR1466800. We acknowledge the science research grants from the China Manned Space Project with Nos. CMS-CSST-2021-A02, CMS-CSST-2021-A03.
The computations in this paper were run on the Gravity Supercomputer at Shanghai Jiao Tong University.

This research used data obtained with the Dark Energy Spectroscopic Instrument (DESI). DESI construction and operations is managed by the Lawrence Berkeley National Laboratory. This material is based upon work supported by the U.S. Department of Energy, Office of Science, Office of High-Energy Physics, under Contract No. DEAC0205CH11231, and by the National Energy Research Scientific Computing Center, a DOE Office of Science User Facility under the same contract. Additional support for DESI was provided by the U.S. National Science Foundation (NSF), Division of Astronomical Sciences under Contract No. AST-0950945 to the NSF’s National Optical-Infrared Astronomy Research Laboratory; the Science and Technology Facilities Council of the United Kingdom; the Gordon and Betty Moore Foundation; the HeisingSimons Foundation; the French Alternative Energies and Atomic Energy Commission (CEA); the National Council of Science and Technology of Mexico (CONACYT); the Ministry of Science and Innovation of Spain (MICINN), and by the DESI Member Institutions: www.desi.lbl.gov/collaborating-institutions.

The DESI Legacy Imaging Surveys consist of three individual and complementary projects: the Dark Energy Camera Legacy Survey (DECaLS), the Beijing-Arizona Sky Survey (BASS), and the Mayall z-band Legacy Survey (MzLS). DECaLS, BASS and MzLS together include data obtained, respectively, at the Blanco telescope, Cerro Tololo Inter-American Observatory, NSF’s NOIRLab; the Bok telescope, Steward Observatory, University of Arizona; and the Mayall telescope, Kitt Peak National Observatory, NOIRLab. NOIRLab is operated by the Association of Universities for Research in Astronomy (AURA) under a cooperative agreement with the National Science Foundation. Pipeline processing and analyses of the data were supported by NOIRLab and the Lawrence Berkeley National Laboratory (LBNL). Legacy Surveys also uses data products from the Near-Earth Object Wide-field Infrared Survey Explorer (NEOWISE), a project of the Jet Propulsion Laboratory/California Institute of Technology, funded by the National Aeronautics and Space Administration. Legacy Surveys was supported by: the Director, Office of Science, Office of High Energy Physics of the U.S. Department of Energy; the National Energy Research Scientific Computing Center, a DOE Office of Science User Facility; the U.S. National Science Foundation, Division of Astronomical Sciences; the National Astronomical Observatories of China, the Chinese Academy of Sciences and the Chinese National Natural Science Foundation. LBNL is managed by the Regents of the University of California under contract to the U.S. Department of Energy. The complete acknowledgments can be found at https://www.legacysurvey.org/acknowledgment/.

We additionally made use of Astropy, a community-developed core Python package for Astronomy (Astropy Collaboration et al. 2018), IPython (Pérez \& Granger 2007), Matplotlib (Hunter 2007), and TOPCAT (Taylor 2005, http://www.starlink.ac.uk/topcat/
).


\begin{appendix}
\setcounter{figure}{0}
\setcounter{table}{0}
\renewcommand{\thefigure}{\Alph{figure}}
\renewcommand{\thetable}{\Alph{table}}

\section{Cosmic variance in the SV3-BGS sub-sample}
\label{appendix:A}
 
\begin{figure*}[!htb]
\centering
\includegraphics[width=0.9\textwidth]{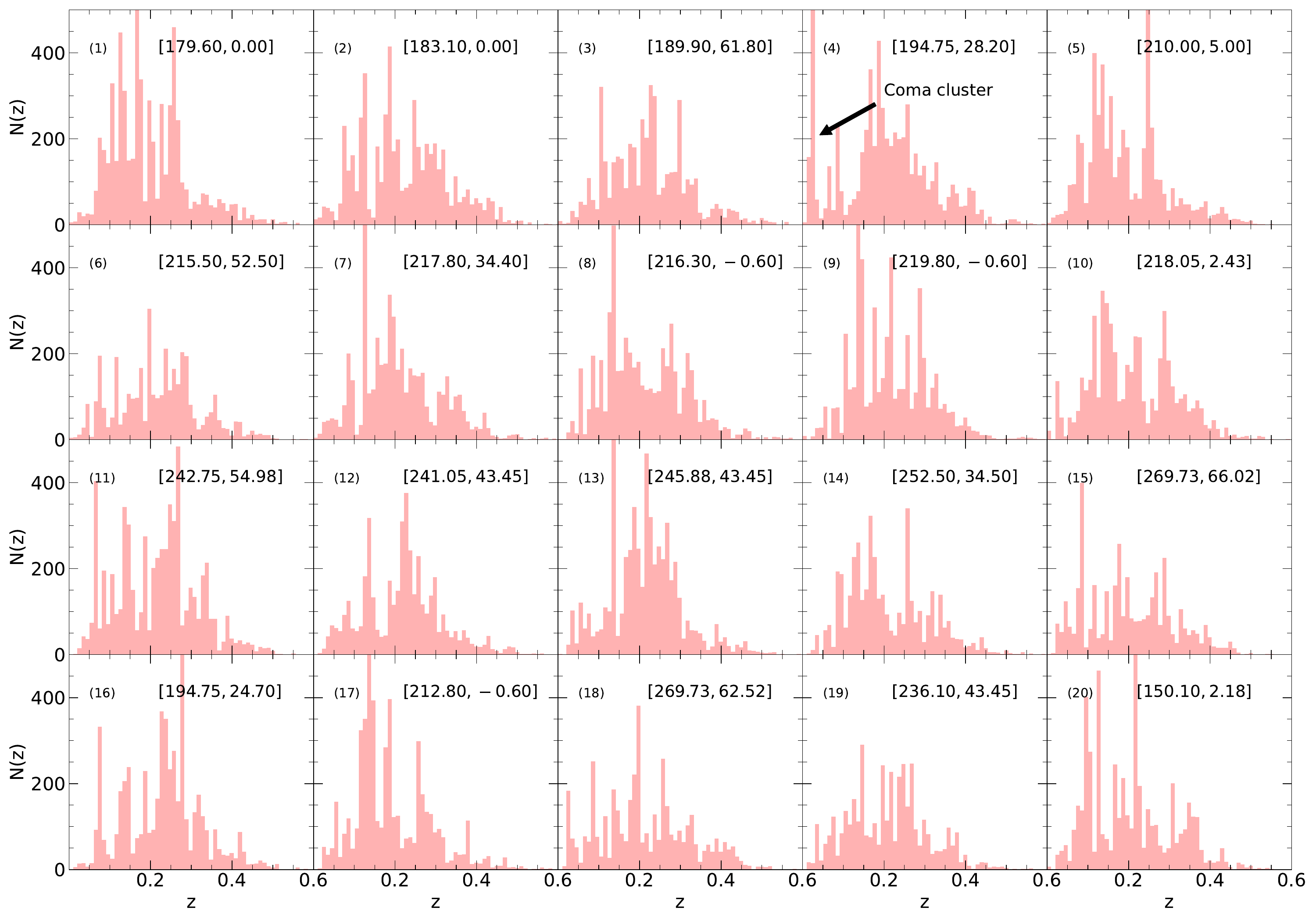}
\caption{Redshift distribution of galaxies in 20 rosettes from SV3 ($1\%$ sky coverage) within $r$-band magnitude 19.5 cut. The fourth rosette at the coordinate position [194.75, 28.20] contains the Coma cluster. \label{fig:20circle}}
\end{figure*}

When performing the LF measurements from the SV3-BGS sub-samples using the method outlined in Section \ref{sec:LF&SMF}, we find a significant enhancement at $L \sim 10^{8}\Lsunhh$ in the low redshift bin compared with that of Y1-BGS sub-sample. To find the cause of the big enhancement, we checked the redshift distribution of galaxies in the 20 rosettes of the SV3-BGS sub-sample. The results are shown in different panels of Figure \ref{fig:20circle}, each corresponding to a particular rosette.  In several panels, spikes exhibit at low redshift, especially in panel (4) which is centered at the coordinate [RA194.75, DEC28.20]. According to the DESI official website\footnote{\url{https://desi.lbl.gov/trac/wiki/SurveyOps/OnePercent}}, this fourth rosette contains the Coma cluster with redshift $\sim$0.0231. 

\begin{figure*}[!htb]
\centering
\includegraphics[width=0.85\textwidth]{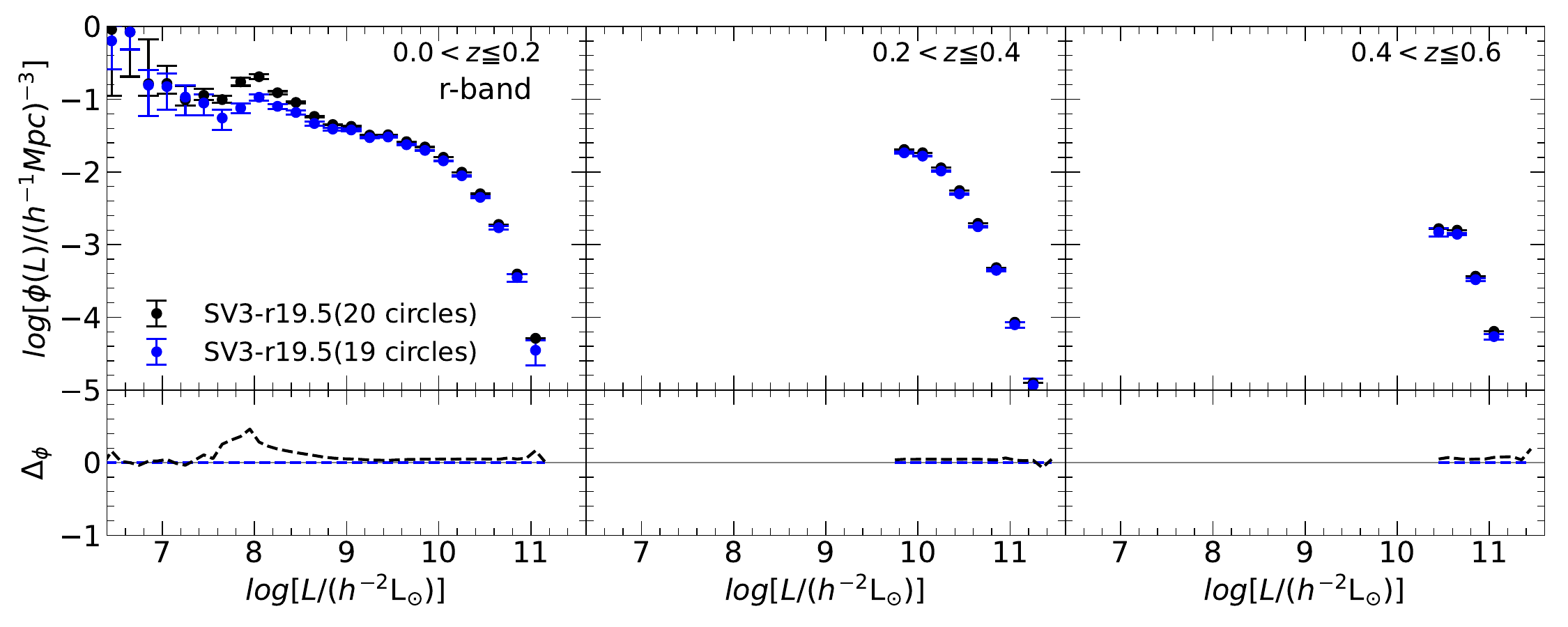}
\includegraphics[width=0.85\textwidth]{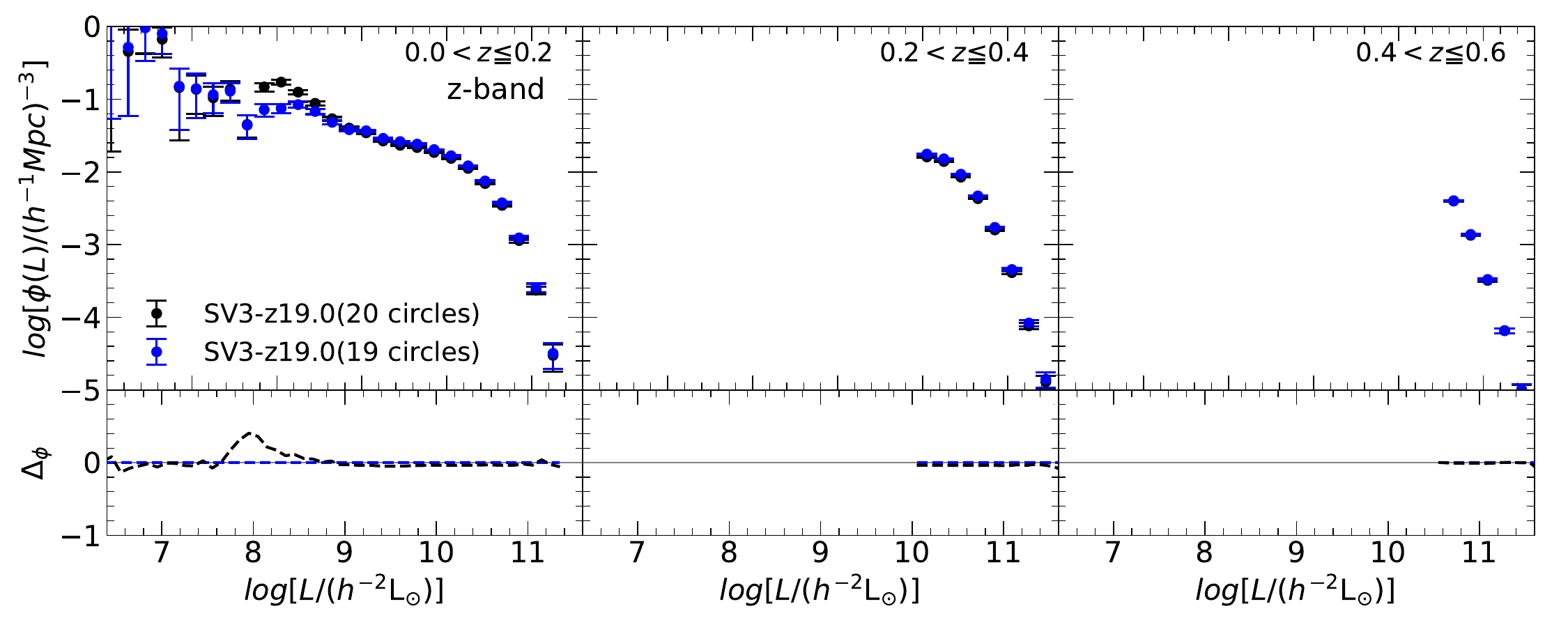}
\caption{Comparison of LFs obtained from 20 and 19 (exclude the fourth) rosettes in the SV3-BGS sub-samples.\label{fig:19circle_LF}}
\end{figure*}

The presence of the Coma cluster significantly enhanced the LF measurements of the SV3-BGS sub-sample at $L \sim 10^{8}\Lsunhh$, which is verified in Figure \ref{fig:19circle_LF}. The black dots with error bars stand for the results with 20 rosettes, and the blue ones are the measurements with 19 rosettes excluding Coma cluster one. Since the number of large clusters in the local universe below the redshift of 0.03 is small, the Coma cluster causes a considerable cosmic variance in the DESI SV3-BGS 1\% sky coverage. Thus, in our investigation, the galaxies in rosette 4 are excluded from our SV3-BGS sub-sample.

\section{Spectroscopic and photometric redshift distribution of SV3-BGS galaxies}
\label{appendix:B}

\begin{figure*}[!htb]
\centering
\includegraphics[width=0.9\textwidth]{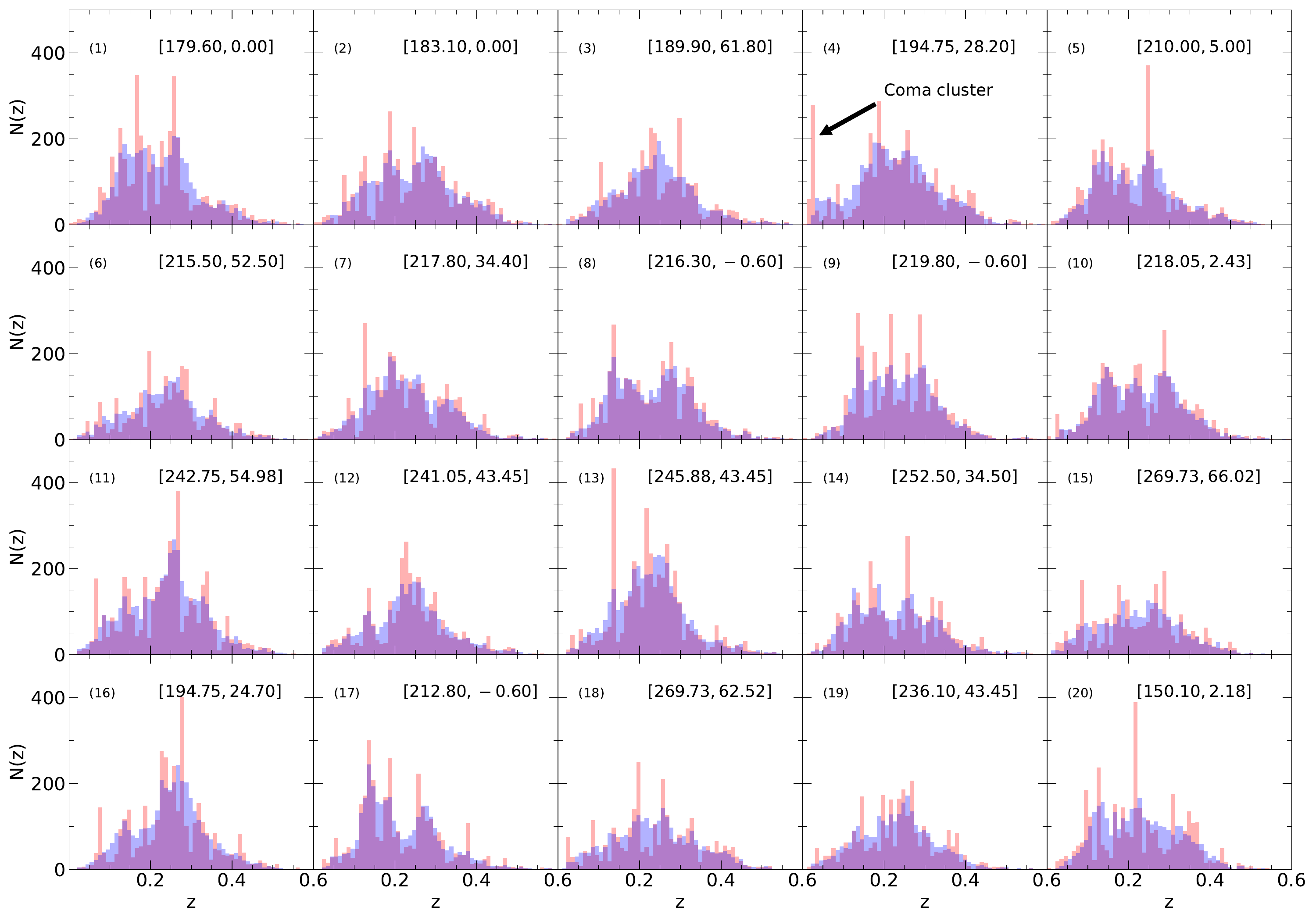}
\caption{Redshift distribution of galaxies in 20 rosettes from SV3 ($1\%$ sky coverage) within $r$-band magnitude  18.5 $\sim$ 19.5 cut. Red and blue histograms show the spec-z and photo-z distributions, respectively. \label{fig:zdist2}}
\end{figure*}

As illustrated in section \ref{sec:data_Jiutian}, the Gaussian distribution of photo-z fails to mimic the diminishing trend of LFs at the faint end within the lowest redshift bin. Consequently, we investigate the actual photo-z distribution for galaxies in the SV3-BGS sample. Displayed in each panel of Figure \ref{fig:zdist2} are the observed spectroscopic and photometric redshift distributions for SV3-BGS galaxies with an apparent magnitude of $18.5<r<19.5$. The photo-z for all galaxies were provided by \citet{Zhou2021}. In comparison to the spectroscopic redshifts, the photo-z distribution at the lowest spectroscopic redshift peak in each panel tends to shift towards higher redshifts.

\end{appendix}

\clearpage
\bibliographystyle{aasjournal}
\bibliography{main}

\end{CJK*}
\end{document}